\documentclass[twocolumn,pra,aps,nofootinbib]{revtex4-1}
\usepackage{amsmath}
\usepackage{amssymb}
\usepackage{graphicx}
\usepackage{comment}
\usepackage{nicefrac}
\usepackage{caption}
\usepackage{subcaption}
\usepackage{hyperref}
\hypersetup{
  colorlinks   = false, 
  urlcolor     = blue, 
  linkcolor    = blue, 
  citecolor   = blue
}
\usepackage{multirow}
\usepackage{tikz, pgfplots}
\usepackage{braket}
\usepackage[title]{appendix}
\usepackage{soul}
\usepackage{standalone}
\usepackage{stmaryrd}
\usepackage[title]{appendix}

\providecommand{\cnot}{\textsc{cnot }{}}

\providecommand{\qec}[3]{$\left \llbracket #1,\,#2,\,#3 \right \rrbracket$}

\begin{document}
\title{Fault-tolerant Quantum Error Correction on Near-term Quantum Processors using Flag and Bridge Qubits}
\author{Lingling Lao, Carmen G. Almudever}
\affiliation{QuTech, Delft University of Technology, The Netherlands}

\begin{abstract}
Fault-tolerant (FT) computation by using quantum error correction (QEC) is essential for realizing large-scale quantum algorithms.
Devices are expected to have enough qubits to demonstrate aspects of fault tolerance in the near future.
However, these near-term quantum processors will only contain a small amount of noisy qubits and allow limited qubit connectivity.
Fault-tolerant schemes that not only have low qubit overhead but also comply with geometrical interaction constraints are therefore necessary.
In this work, we combine flag fault tolerance with quantum circuit mapping, to enable an efficient \emph{flag-bridge} approach to implement FT QEC on near-term devices.
We further show an example of performing the Steane code error correction on two current superconducting processors and numerically analyze their performance with circuit level noise.
The simulation results show that the QEC circuits that measure more stabilisers in parallel have lower logical error rates.
We also observe that the Steane code can outperform the distance-3 surface code using flag-bridge error correction.
In addition, we foresee potential applications of the flag-bridge approach such as FT computation using lattice surgery and code deformation techniques.
\end{abstract}

\maketitle

\section{Introduction}
Near-term quantum processors will consist of fifty to a few hundred noisy qubits and allow a limited number of faulty gates. 
They are also known as Noisy-Intermediate-Scale Quantum (NISQ)~\cite{preskill2018quantum} processors.
For instance, Google, IBM, and Intel have respectively announced 72-qubit~\cite{google72qubit}, 50-qubit~\cite{ibm50qubit}, and 49-qubit~\cite{intel49qubit} superconducting processors which have coherence times of $\sim$ 100 microseconds and two-qubit gate error rates near $0.1\%$~\cite{barends2014superconducting}.
Many efforts have been focusing on designing special quantum applications~\cite{hempel2018quantum, kokail2019self} and developing compilation techniques~\cite{fu2016heterogeneous, fu2017experimental} such that one can solve practical problems and even demonstrate quantum supremacy on NISQ processors only using noisy bare qubits. 

However, fault tolerance will be necessary to reliably implement large-scale quantum algorithms.
This can be achieved through the use of active quantum error correction (QEC).
The idea of QEC is to encode one logical qubit into many physical qubits and repeatedly perform syndrome extraction to detect and correct errors.
Both the encoding and error detection procedure should be fault-tolerant (FT).
Furthermore, operations on these logical qubits need to be performed fault-tolerantly.
Although the high qubit overhead of QEC makes it difficult to realize scalable FT computation in the near future, we can begin to show how fault tolerance works in practice.
The first step is to demonstrate fault-tolerant quantum error correction, that is, FT quantum memory.

General fault-tolerant quantum error correction protocols such as those from Shor~\cite{shor1996fault}, Steane~\cite{steane1997active}, and Knill~\cite{knill2005scalable} can be applied to various stabiliser codes. 
However, these error correction schemes all require many ancilla qubits, which are scarce resources in near-term quantum processors.
In order to perform FT QEC with low qubit overhead, a new error correction protocol has been proposed~\cite{yoder2017surface,chao2018quantum,chamberland2018flag,reichardt2018fault}.
It replaces a non-FT syndrome extraction circuit by a circuit which can detect correlated (or hook) errors by adding only one or a few extra ancilla qubits, called \textbf{flag} qubits. 


This flag QEC scheme provides an efficient way to demonstrate fault tolerance in small experiments.
However, many orthodox flag circuits couple one qubit to many others, requiring high-degree qubit connectivity.
It is difficult or even impossible to directly map available flag circuits onto near-term quantum processors which have geometrical interaction constraints such as the nearest-neighbour connectivity in superconducting processors~\cite{ibm17experience, versluis2017scalable, rigetti}. 
One may need to apply extra operations such as SWAP gates to move qubits to be adjacent, increasing the circuit size in terms of depth and total gate number, or even circuit width.
More importantly, the resulting circuit may not be fault-tolerant, or produce higher error rates when used.

In this work, we extend the set of available flag circuits to a variety of equivalent circuits that can perform the same stabiliser measurement fault-tolerantly.
In these circuits, the flag qubits are also used as \textbf{bridges} to cope with the connectivity constraints, called \textbf{flag-bridge} qubits.
Using these circuits, one can fault-tolerantly map a QEC code to a given processor with low overhead by choosing appropriate flag-bridge circuits.
We also develop a simulation framework to automate the procedure of fault tolerance checking, decoder design (including a look-up-table decoder and a neural-network decoder) for given flag-bridge circuits of some low-distance QEC codes.
This automation is desirable for demonstrating fault-tolerant quantum error correction in small experiments.
Moreover, we present mapping examples of the Steane code on two different qubit processor topologies and analyze their fault tolerance numerically.
In addition, we show the proposed flag-bridge approach can be applied to FT computation implemented by lattice surgery and code deformation techniques.

The rest of this paper is organized as follows.
We first review the basics of flag-based quantum error correction in Section~\ref{sec:flag}.
Then we introduce the proposed flag-bridge approach in Section~\ref{sec:bridge}.
Afterwards, the mapping of the Steane code onto two qubit processor topologies and corresponding numerical results are shown in Section~\ref{sec:steane}.
Moreover, we provide the potential applications of flag-bridge circuits in Section~\ref{sec:applications}. 
Finally, Section~\ref{sec:conclude} concludes the paper.

\section{Flag-based quantum error correction}
\label{sec:flag}
In this section, we briefly introduce the flag-based error syndrome extraction for stabiliser codes. For more details, we refer the readers to~\cite{yoder2017surface, chao2018quantum, chamberland2018flag, reichardt2018fault, divincenzo2007effective}. 

\begin{figure}[htb!]
 \centering
    \begin{subfigure}[b!]{0.24\textwidth}
    \includegraphics[width=1\textwidth]{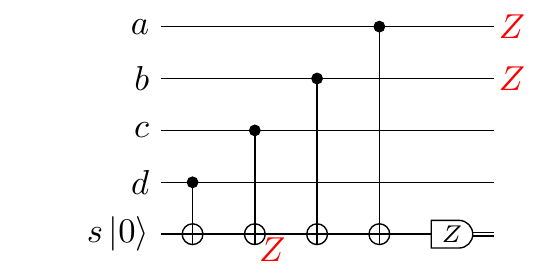}
    \caption{}
    \label{fig:checkz_1a}
    \end{subfigure}
 \begin{subfigure}[b!]{0.25\textwidth}
    \includegraphics[width=1\textwidth]{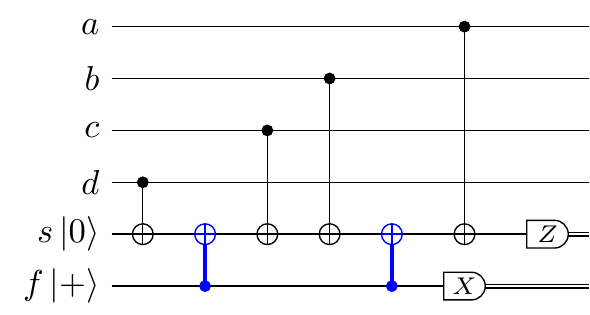}
    \caption{}
    \label{fig:checkz_2a1}
    \end{subfigure}
    \begin{subfigure}[b!]{0.21\textwidth}
    \includegraphics[width=1\textwidth]{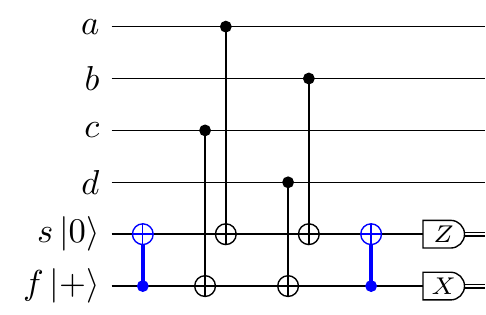}
    \caption{}
    \label{fig:checkz_2a2}
    \end{subfigure}
\caption{The syndrome extraction circuits for the $Z_{a,b,c,d}$  operator, where $s$ is the syndrome qubit and $f$ is the flag qubit. (a) The circuit only using one syndrome ancilla may not be fault-tolerant. For example, one fault ($Z_{s}$) on the second \cnot gate could lead to correlated weight-2 errors on data qubits ($Z_{a}, Z_{b}$), which may not be correctable. (b) and (c) The flag-based circuits can detect these hook errors~\cite{yoder2017surface,chao2018quantum,chamberland2018flag}.}
\label{fig:checkz}
\end{figure}

Figure~\ref{fig:checkz} shows the circuits for measuring a weight-$4$ $Z$-stabiliser (or check), similar circuits can be derived for measuring other Pauli operators.
In all the circuits presented in this paper, a \cnot gate between a data qubit and an ancilla qubit is called an s-\cnot (in black) and a \cnot gate between two ancilla qubits is called an f-\cnot (in blue).
Generally, the syndrome for this $Z$-check can be extracted using the circuit with only one ancilla qubit (Figure~\ref{fig:checkz_1a}).
However, this circuit is not fault-tolerant because one single fault could cause 2 or more data errors.
These correlated errors may lead to failures of some QEC codes.
The surface code is an exception which can correct these hook errors if the two-qubit gates are performed in a specific order~\cite{fowler2012surface}.
In order to perform fault-tolerant quantum error correction, one can use the flag circuits in Figures~\ref{fig:checkz_2a1} and~\ref{fig:checkz_2a2} that only add one extra ancilla qubit.
When there is no fault, each of these flag circuits behaves the same as the non-FT one.
When there is a fault that can lead to hook errors, the measurement of the flag qubit will be nontrivial such that the hook errors are detected.
For instance, if the same fault in Figure~\ref{fig:checkz_1a} happens in the circuit of Figure~\ref{fig:checkz_2a1}, then the measurement of qubit $f$ will be `1' (raising a flag).

\begin{figure}[htb!]
 \centering
 \includegraphics[width=0.4\textwidth]{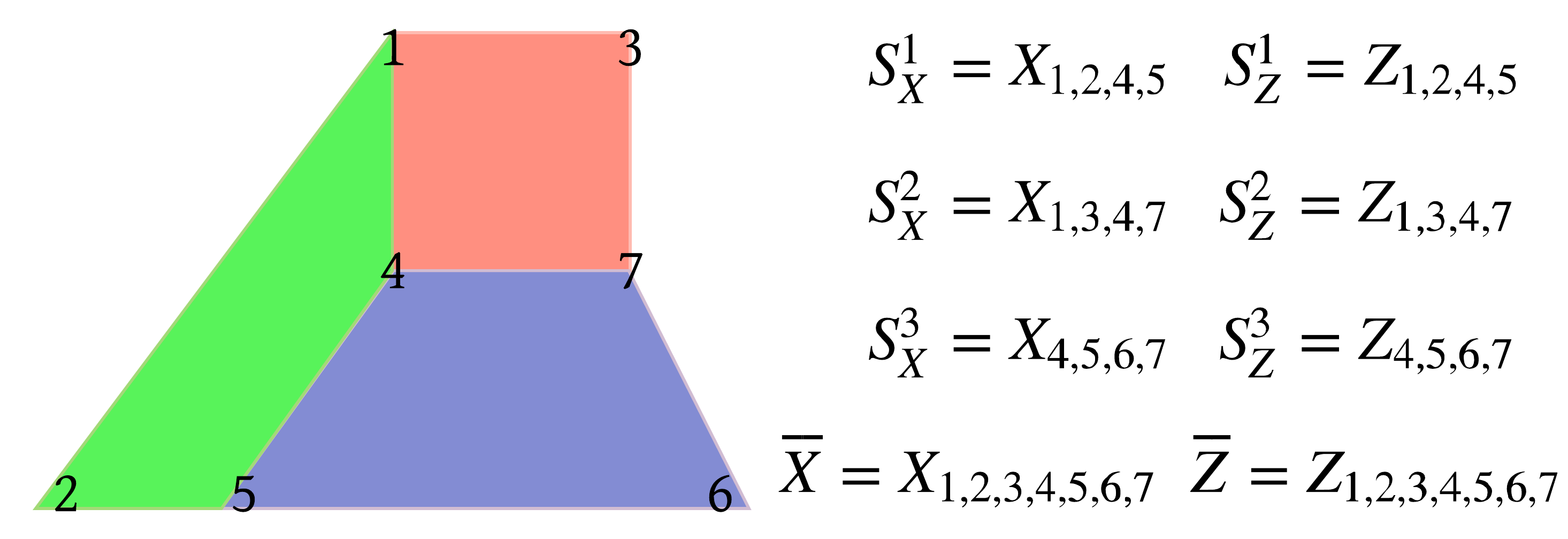}
\caption{(Left) The qubit layout of the \qec{7}{1}{3} Steane code. Data qubits are on the vertices and each plaquette represents two stabilisers: one weight-$4$ $X$-stabiliser and one weight-$4$ $Z$-stabiliser. (Right) all the six stabiliser generators and logical $X$ and $Z$ operators.}
\label{fig:steaneqec}
\end{figure}

\begin{figure}[htb!]
 \centering
     \begin{subfigure}[b!]{0.4\textwidth}
    \includegraphics[width=0.8\textwidth]{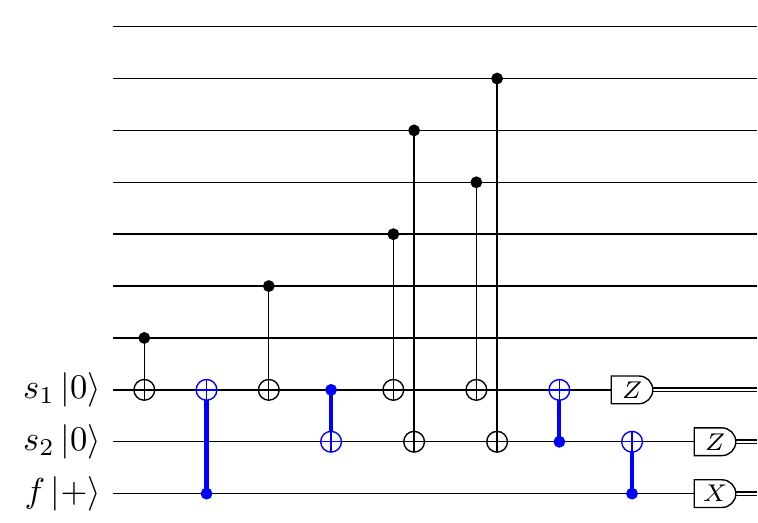}
    \caption{}
    \label{fig:check44_3a_42}
    \end{subfigure}
  \begin{subfigure}[b!]{0.4\textwidth}
    \includegraphics[width=\textwidth]{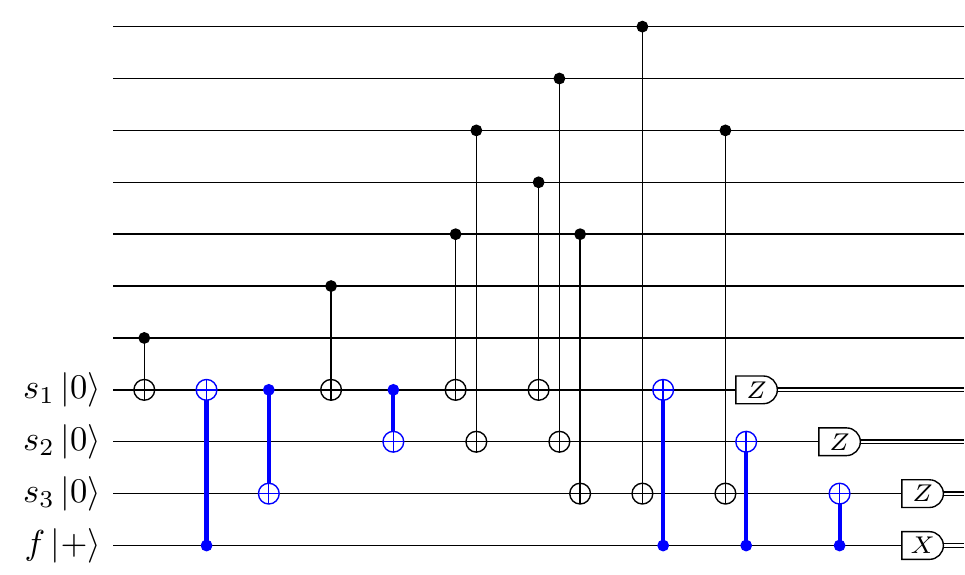}
    \caption{}
    \label{fig:check444_4a_432}
    \end{subfigure}
\caption{Flag circuits~\cite{chao2018quantum,reichardt2018fault} for (a) measuring two weight-$4$ $Z$-checks in parallel using three ancillas and (b) measuring three weight-$4$ $Z$-checks in parallel using four ancillas.}
\label{fig:steane_multicheck1}
\end{figure}

Flag-based quantum error correction can be applied to many codes such as the \qec{5}{1}{3} code, Hamming codes, surface codes, color codes, etc.
For example, fault-tolerant QEC for the smallest color code, the Steane code in Figure~\ref{fig:steaneqec}, can be realized as follows: 
first measure each stabiliser generator one by one using flag circuits similar to those in Figure~\ref{fig:checkz}; if a flag raises or a syndrome appears, then stop this round\footnote{A full round of error syndrome extraction is defined as measuring all the stabiliser generators of the code for one time.} and sequentially measure all the stabilisers using the non-FT syndrome extraction circuit.
Note that if connectivity is fixed, we can't necessarily change the syndrome measurement circuit all of a sudden.
One can use only two ancilla qubits to perform FT QEC for the Steane code at the cost of using more time steps.
However, many quantum systems have very short coherence times~\cite{riste2015detecting, kelly2015state, ibm17experience}.
Parallelizing stabiliser measurement will be beneficial to achieve lower logical error rates.
Chao and Reichardt~\cite{chao2018quantum, reichardt2018fault} have proposed several circuits to perform two or three parity checks in parallel for the Steane code.
The circuits they propose for measuring two and three $Z$-checks at the same time using only one flag qubit are shown in Figure~\ref{fig:steane_multicheck1}.
As shown, more ancilla qubits are required to achieve this parallelism compared to the sequential stabiliser measurement circuits.
This implies there is a trade-off between the number of qubits required and the number of stabilisers that can be measured simultaneously.

Flag-based syndrome extraction is promising for demonstrating quantum error correction and fault tolerance in small quantum experiments because of its low qubit overhead. 
However, current or near-term quantum processors have many hardware limitations.
One of the main constraints is the degree of qubit connectivity, that is, one qubit can only interact with a limited number of other qubits.
It is challenging to map existing flag circuits onto connectivity-constrained quantum processors meanwhile maintaining the fault tolerance with low costs.
For instance, the ancilla qubit $s$ of the flag circuit in Figure~\ref{fig:checkz_2a1} needs to interact with five qubits, which cannot be supported in a grid topology where each qubit only has at most four neighbours~ such as the one in \cite{versluis2017scalable}.
Besides, general circuit mapping techniques~\cite{lao2018mapping,li2019tackling,tannu2019not,shi2019optimized,lao2019mapping} that move qubits to be adjacent by applying SWAP gates will lead to high overhead in the circuit size. 
More importantly, it may result in higher logical error rates or even destroy the fault tolerance of the QEC circuits because of the error propagation through two-qubit gates.
In this work, we propose a flag-bridge approach to solve this mapping problem, which will be explained in the next section.

\section{Flag-bridge quantum error correction}
\label{sec:bridge}

In this section, we illustrate the proposed flag-bridge approach which allows fault-tolerant quantum error correction with low qubit overhead on connectivity-limited quantum processors.
\subsection{Flag-bridge syndrome extraction circuits}
\begin{figure}[tbh!]
 \centering
     \begin{subfigure}[h]{0.24\textwidth}
     \includegraphics[width=\textwidth]{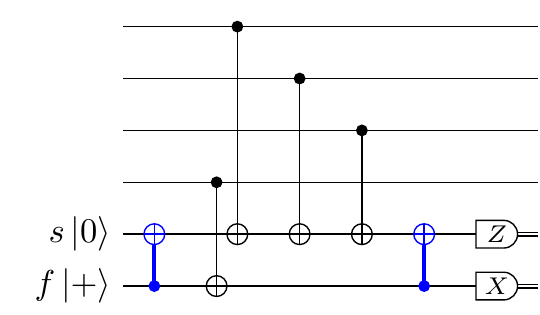}
     \caption{}
    \label{fig:check4_2a_13}
    \end{subfigure}
    \begin{subfigure}[h]{0.24\textwidth}
    \includegraphics[width=\textwidth]{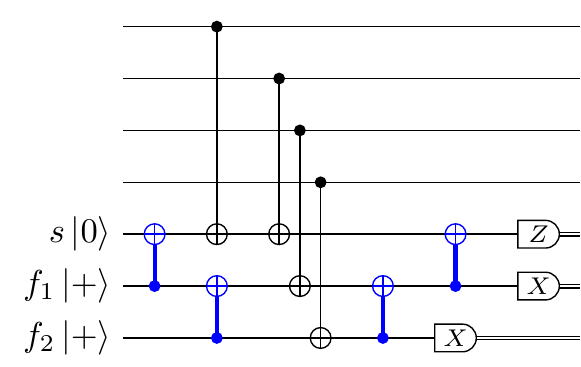}
    \caption{}
    \label{fig:check4_3a_121}
    \end{subfigure}
\caption{Flag-bridge circuits for measuring one weight-$4$ $Z$-check using (a) two ancillas and (b) three ancillas. 
}
\label{fig:steane1check}
\end{figure}

We first provide a microscopic explanation of how a flag-based circuit can perform a specific stabiliser measurement using the stabiliser formalism~\cite{gottesman1998heisenberg}.
Then we generalise this flag scheme such that one can extend available flag circuits to more equivalent ones that are different in terms of the total number of gates, circuit depth, and connectivity requirement. 
We will use the circuit in Figure~\ref{fig:checkz_2a2} as an example.
A flag syndrome extraction circuit can be understood as a circuit that replaces the bare ancilla qubit by an `encoded' ancilla up to gate commutation.
As shown in Figure~\ref{fig:checkz_2a2}, the first blue \cnot gate entangles ancilla qubit $s$ and qubit $f$ (the encoding circuit), encoding a logical ancilla in a \qec{2}{1}{1} error detection code of which stabiliser is 
\[\left \langle X_{s} \otimes X_{f}\right \rangle\]
and logical operators are 
\[\left \langle \overline{X}=X_{s}, \overline{Z}=Z_{s} \otimes Z_{f}\right \rangle.\]
This logical qubit is fixed in the $\overline{Z}$ basis.
Then one can perform stabiliser measurement using this logical ancilla.
Assume the four data qubits ($a,b,c,d$) are initially stabilized by $(-1)^{y}Z_{a,b,c,d}$, the four subsequent s-\cnot gates between data qubits and ancilla qubits will keep the stabilisers of all the qubits invariant, which are, 
\[\left \langle  X_{s} \otimes X_{f}, (-1)^{y}Z_{4,5,6,7}\right \rangle,\]
but it will gradually transform the logical operators into
\[\left \langle  \overline{X}=X_{s}, \overline{Z}=Z_{s} \otimes Z_{f} \otimes (-1)^{y}Z_{4,5,6,7}\right \rangle.\]
More generally, since $X_{f}$ and $X_{s}$ have the same effect on the encoded ancilla state, one can perform each s-\cnot gate between the particular data qubit with any ancilla qubit. 
Specifically, in the encoded ancilla area, $k_{s}$ and $k_{f}$ s-\cnot gates can be applied on ancillas $s$ and $f$ respectively, where $k_{s}$ and $k_{f}$ are integers and $k_{s} + k_{f} = 4$.
For example, the circuit shown in Figure~\ref{fig:check4_2a_13} also performs a weight-4 $Z-$stabiliser measurement equivalent to this circuit (Figure~\ref{fig:checkz_2a2}), where $k_{s}=3$ and $k_{f}=1$. 

Afterwards, the last f-\cnot (the decoding circuit) disentangles these two ancillas, leading to the final stabiliser to be
\[ \left \langle (-1)^{y}Z_{4,5,6,7}\right \rangle\]
and the logical operators of these ancillas to be
\[ \left \langle X_{f},  (-1)^{y}Z_{s}\right \rangle.\]
This means the readout $y$ of measurement $M_{z}$ on ancilla $s$ indicates the measurement result of the stabiliser $Z_{a,b,c,d}$.
Therefore, this circuit indeed measures a weight-$4$ $Z$-check.
Besides, the measurement result of ancilla $f$ implies the syndrome of the \qec{2}{1}{1} code, that is, it can detect one single $Z$ error that occurs on any ancilla and then raises a flag.

Once a flag circuit based on the above approach is generated, one can transform it into other equivalent ones that can perform the same stabiliser measurement by applying gate commutation, e.g., the circuit in Figure~\ref{fig:checkz_2a1}.
Note that the circuits generated by commuting gates may not be fault-tolerant.

Moreover, one can use a larger `encoded' ancilla to measure a weight-$n$ $Z$-check (similar circuits can be applied to other Pauli operators).
This logical ancilla is encoded by $m$ physical qubits denoted by a set $Q=\{1, 2, \cdots, m\}$, where one is syndrome qubit ($Q_{s}=\{1\}$) and the other $m-1$ are flag qubits ($Q_{f}=\{2, \cdots, m\}$). The underlying error detection code \qec{m}{1}{1} of this logical ancilla has stabilisers
\[ \left \langle X_{j}\otimes X_{k}\right \rangle, j\in Q_{s},k \in Q_{f}\]
and logical operators
\[ \left \langle \overline{X}=X_{j}, \overline{Z}=\bigotimes_{i} Z_{i} \right \rangle, i, j\in Q.\]
Similar to the two-ancilla flag circuits, this weight-$n$ check can be distributed to all $m$ ancillas, $k_{i}$ s-\cnot gates will be applied on ancilla $i$, where $\sum_{i=1}^{m}k_{i} = n$.
For example, the circuit in Figure~\ref{fig:check4_3a_121} measures one weight-4 $Z$-stabiliser using one syndrome qubit ($s$) and two flag qubits ($f_{1}, f_{2}$), each qubit only needs to interact with at most three others.

In addition, one can also measure $p$ $Z$-checks in parallel by encoding $p$ logical ancillas into $m$ physical ancillas.
The underlying \qec{m}{p}{1} code is stabilized by 
\[ \left \langle X_{i} \otimes \bigotimes_{j} X_{j} \right \rangle, i\in Q_{f}, j \in Q_{s}.\]
Its $p$ logical operators are
\[ \left \langle \overline{X_{k}}=X_{i},  \overline{Z_{k}}=Z_{i} \otimes Z_{j}\right \rangle, i,j\in Q, i< j.\]
Where, $Q_{s}$ is the set of $p$ syndrome qubits and $Q_{f}$ is the set of $m-p$ flag qubits.
After the encoding of ancilla qubits, one can simply assign all the s-\cnot gates for performing one check to a particular syndrome qubit.
Furthermore, one can reduce the total number of s-\cnot gates by applying gate commutation when two or more checks are performed on the same data qubit(s).
Figure~\ref{fig:check44_3a_42} and Figure~\ref{fig:check444_4a_432} show the flag circuits to measure two and three checks of the Steane code in parallel by using ancillas encoded in a \qec{3}{2}{1} code and in a \qec{4}{3}{1} code, respectively,
These circuits use less s-\cnot gates than required by commuting some \cnot gates out of the encoded area (generally 4 s-\cnot gates are needed for each weight-4 check).

\begin{figure}[tbh!]
 \centering
    \begin{subfigure}[h]{0.4\textwidth}
    \includegraphics[width=0.75\textwidth]{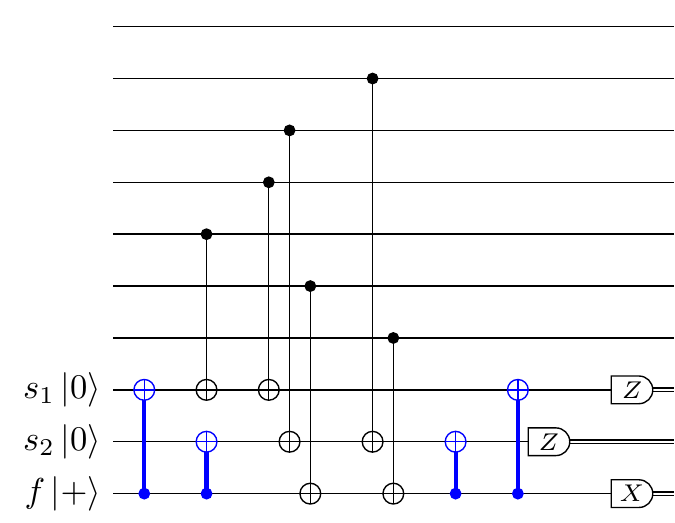}
    \caption{}
    \label{fig:check44_3a_222}
    \end{subfigure}
 \begin{subfigure}[h]{0.4\textwidth}
    \includegraphics[width=1\textwidth]{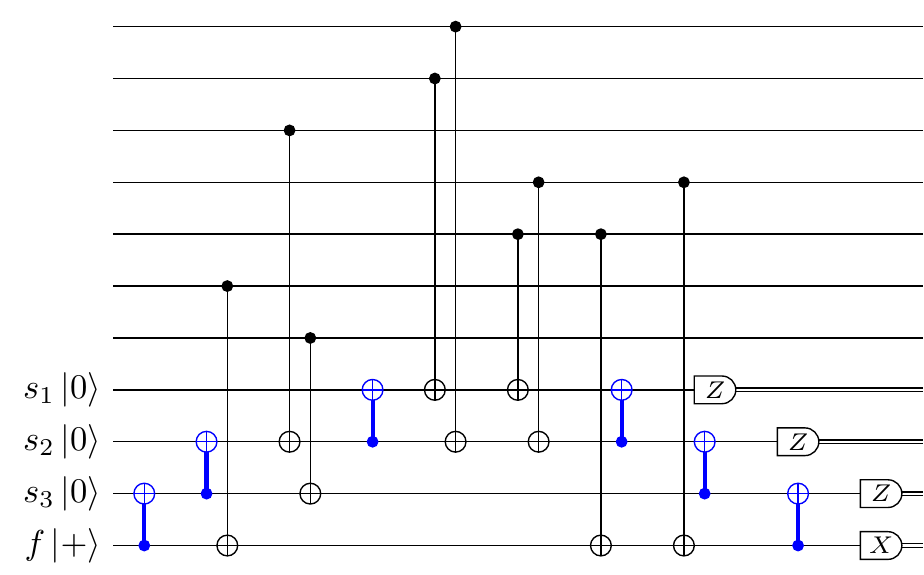}
    \caption{}
    \label{fig:check444_4a_2313}
    \end{subfigure}
\caption{Flag-bridge circuits which measure (a) two and (b) three weight-$4$ $Z$-checks in parallel.}
\label{fig:steane_multicheck2}
\end{figure}

Moreover, one can achieve this gate reduction by distributing some s-\cnot gates to flag qubits, which can even help to reduce the circuit depth.
Figure~\ref{fig:check44_3a_222} shows the example circuit that measures two checks of the Steane code in parallel but uses less timesteps than Figure~\ref{fig:check44_3a_42}.
Note that the s-\cnot distribution for parallel syndrome measurement needs to be designed carefully since one flag qubit is used for flagging multiple checks.
This distribution also depends on the decoding procedure of the \qec{m}{p}{1} code.
Figure~\ref{fig:check444_4a_2313} shows the example circuit that measures three checks using less timesteps than Figure~\ref{fig:check444_4a_432}. 
Besides, the circuits in Figure~\ref{fig:steane_multicheck2} require less degree of qubit connectivity than the ones in Figure~\ref{fig:steane_multicheck1}.

By employing the ideas of encoding ancillas, distributing s-\cnot, and commuting gates, we can generate more equivalent syndrome extraction circuits that have different connectivity requirements.
Note that not all the equivalent circuits generated using this approach are fault-tolerant.
The fault tolerance can be checked based on the error correction protocol, which will be explained in the next section.
For these FT circuits, ancillas are not only used as syndrome and flag qubits to detect errors, but also as bridges to allow the interaction between data qubits and the encoded ancilla block.
Such a syndrome extraction circuit is called a flag-bridge circuit.

\subsection{Fault-tolerant protocol for flag-bridge error correction}

\subsubsection{FT QEC condition}
For distance-3 codes, a QEC circuit is fault-tolerant if it can either immediately correct all errors from a single fault or only leave a weight-1 error to the next cycle. 
A formal condition of FT flag-bridge quantum error correction for distance-3 codes, similar to the flag error correction in~\cite{chamberland2018flag}, can be defined as follows:

Consider a stabiliser code $\mathcal{S}=\left \langle g_{1}, g_{2}, \cdots , g_{r} \right \rangle$ and its QEC circuit $\mathcal{C}$ which is composed of the flag-bridge circuits for measuring the stabiliser generators, that is, $\mathcal{C} =\left \{ c(g_{1}), c(g_{2}),\cdots , c(g_{r}) \right \}$, where $c(g_{i})$ is the flag circuit of measuring stabiliser $g_{i}$. 
Note that the total number of flag-bridge circuits is smaller than $r$ if several stabilisers are measured simultaneously in one flag-bridge circuit.
For all generators $g$, all pairs of elements $E, {E}'\in \mathcal{E}(g)$ satisfy $sf(E) \neq sf({E}')$ or $E \sim {E}'$, where $\mathcal{E}(g)$ is the set of all errors caused by one fault, $sf(E)$ is the syndrome and flag string caused by $E$. 
We define $E \sim {E}'$ to mean that there is an element $g$ in $\mathcal{S}$ such that ${E}'\propto gE$, that is, these errors are stabiliser-equivalent.


Based on this criterion, we check the fault tolerance of each generated QEC circuit $\mathcal{C}$ through a brute-force simulation under circuit level noise, analogous to~\cite{yoder2017surface}.
It is implemented by injecting each individual fault from a circuit-based error model on every single-qubit or two-qubit gate in a given QEC circuit and then collecting the final syndromes and flags.
If there are two or more sets of errors which lead to the same syndrome-flag string but do not
yield a stabiliser when multiplied, then this QEC circuit is not fault-tolerant.

\subsubsection{FT QEC procedure}
A full cycle of fault-tolerant error correction for distance-3 codes using flag-bridge circuits can be performed as follows:
\begin{enumerate}
    \item For the first round of syndrome extraction, each circuit $c(g_{i}) \in \mathcal{C}$ is sequentially performed. If there are non-trivial flags $f_{i}^{1}$ or non-trivial syndromes $s_{i}^{1}$ of $c(g_{i})$, then this round will be terminated and another full round for all circuits in $\mathcal{C}$ will be performed.
    All the syndromes $s^{2}=\bigcup_{i} s_{i}^2$ and flags $f^{2}=\bigcup_{i} f_{i}^2$ of the second round will be collected.
    \item If $f_{i}^{1}$ is not empty, one can decode using $f_{i}^{1}$ and $s^{2}$ (and $f^{2}$).
    If $f_{i}^{1}$ is empty, but $s_{i}^{1}$ is not empty, one can decode using $s^{2}$ (and $f^{2}$). 
    Otherwise, no corrections are needed.
\end{enumerate}

In this FT QEC procedure, we use flag-bridge circuits for both rounds of syndrome extraction because of the connectivity constraint, which is different from the ones proposed in~\cite{chao2018quantum, chamberland2018flag,reichardt2018fault}, where non-FT syndrome extraction circuits that use only one ancilla are executed for the second round. 

\subsubsection{Error decoders}
Normally, error correction of topological codes like surface codes have special structures for the measured syndromes so that one can use heuristic algorithms to find high-probability errors.
These types of decoders such as the minimum weight perfect matching decoder \cite{fowler2015minimum} and the belief propagation decoder \cite{duclos2010fast} can be applied to the same QEC code with different distances.
However, the flag-bridge error correction circuits of a QEC code for a specific quantum platform are ad hoc.
Different circuits may be chosen based on the qubit topology, leading to different error-syndrome patterns and in turn requiring different decoding strategies.  
It is difficult to design heuristic decoding algorithms that can be applied to various syndrome extraction circuits.
Since flag-bridge circuits are likely to be used for low-distance codes in small experiments, a simple decoding solution is to create a look-up table (LUT) for each QEC circuit.
A LUT decoder can find the most likely Pauli errors from a single fault that leads to the observed syndromes and flags.
LUT decoders can be easily derived from the brute-force checking procedure~\cite{yoder2017surface}.

Another type of decoders are the neural-network (NN) decoders \cite{krastanov2017deep, varsamopoulos2017decoding, baireuther2018machine, ni2018neural}.
They can provide high-speed decoding, be adaptable to different error models, and  be more easily implemented on hardware. 
Moreover, a NN decoder can be developed by training the network using only input-output pairs without any knowledge of the QEC code, making it favorable for flag-bridge circuits.
For example, the inputs of a NN decoder are the observed syndromes and its outputs can be the actual physical errors that have occurred.
The implementation details of the LUT decoder and the NN decoder can be found in Appendix~\ref{app:decoder}.

In this work, we design a simulation framework to automate the procedure of fault tolerance checking, LUT generation, and NN decoder training for given flag-bridge syndrome extraction circuits of the Steane code.
This automation is desirable for demonstrating fault-tolerant quantum error correction in near-term processors which may have different geometrical interaction constraints.

\section{Steane code error correction on two device topologies}
\label{sec:steane}

\begin{figure}[tbh!]
 \centering
     \begin{subfigure}[b]{0.18\textwidth}
    \includegraphics[width=\textwidth]{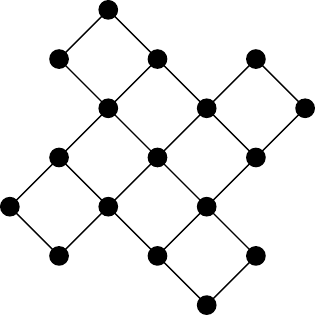}
    \caption{}
    \label{fig:sc_arch}
    \end{subfigure}\hspace{5mm}
      \begin{subfigure}[b]{0.18\textwidth}
    \includegraphics[width=\textwidth]{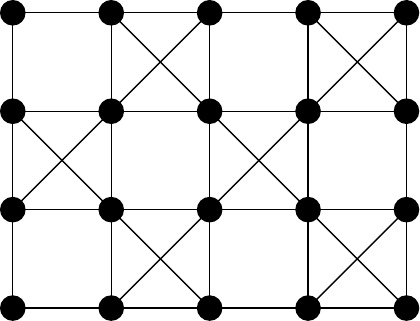}
    \caption{}
    \label{fig:ibm_arch}
    \end{subfigure}
\caption{(a) The Surface-17 topology and (b) the IBM-20 topology, where each node represents a qubit, and each edge indicates the connectivity between two qubits.}
\label{fig:arch}
\end{figure}

In this section, we show how to map the Steane code error correction onto two different processors with limited connectivity using the proposed flag-bridge circuits, namely, the Surface-17 transmon processor (Surface-17)~\cite{versluis2017scalable} and the IBM Q Tokyo processor (IBM-20)~\cite{ibm17experience} (Figure~\ref{fig:arch}).
Furthermore, we numerically analyze each flag-bridge quantum error correction procedure under circuit level noise.
This error model inserts depolarizing errors after each operation in a flag-bridge circuit as follows: 
1) each single-qubit gate is followed by a $X$, $Y$, or $Z$ with probability $\nicefrac{p}{3}$; 2) each two-qubit gate is followed by an element of $\{I,X,Y,Z\}^{\bigotimes 2}\backslash \{II\}$ with probability $\nicefrac{p}{15}$; 3) the preparation or measurement in the $Z$ basis is flipped with probability $p$.
The elementary Clifford operations used in this simulation are preparation and measurement in the $Z$ basis, $H$ and \cnot gates. 
Other operations need to be further decomposed into these elementary operations.
For example, each control-phase gate is replaced by two $H$ gates and one \cnot gate.

\subsection{Mapping}
\begin{figure}[tbh!]
 \centering
     \begin{subfigure}[h]{0.2\textwidth}
    \includegraphics[width=0.9\textwidth]{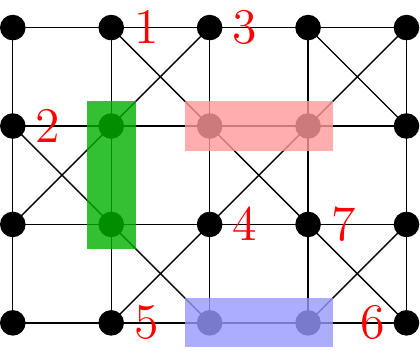}
    \caption{Steane-cl-L2}
    \label{fig:steane2ibm222}
    \end{subfigure}\hspace{5mm}
  \begin{subfigure}[h]{0.2\textwidth}
    \includegraphics[width=0.9\textwidth]{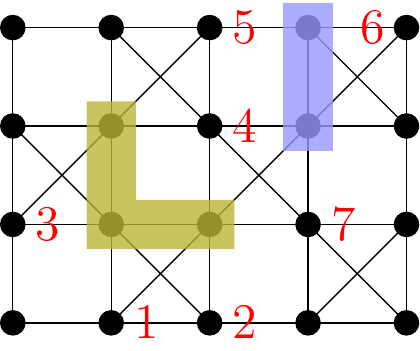}
    \caption{Steane-c2-L2}
    \label{fig:steane2ibm33}
    \end{subfigure}\hspace{5mm}
      \begin{subfigure}[h]{0.2\textwidth}
    \includegraphics[width=0.9\textwidth]{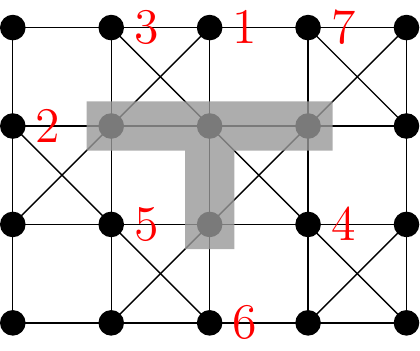}
    \caption{Steane-c3-L2}
    \label{fig:steane2ibm4}
    \end{subfigure}
\caption{Mapping of the Steane code onto the IBM-20 topology (a) using the flag-bridge circuits in Figures~\ref{fig:checkz_2a2} and~\ref{fig:check4_2a_13}; (b) using the flag-bridge circuits in Figures~\ref{fig:check4_3a_121} and~\ref{fig:check44_3a_222}; (c) using the four-ancilla flag-bridge circuit in Figure~\ref{fig:check444_4a_2313} to measure three stabilisers simultaneously.
}
\label{fig:steane2ibm20}
\end{figure}

Many current and NISQ processors have geometrical connectivity constraints, that is, each qubit can only interact with a few neighbours.
It is challenging or even impossible to directly perform existing flag-based quantum error correction without adding more operations and/or without losing fault tolerance.
For example, the flag circuit which measures one weight-4 $Z$-stabiliser of the Steane code in Figure~\ref{fig:checkz_2a1} cannot be directly executed on the Surface-17 topology (Figure~\ref{fig:sc_arch}) but can be supported by the IBM-20 topology (Figure~\ref{fig:ibm_arch}).
This is because qubit $s$ needs to interact with 5 qubits but in Surface-17 each qubit has at most 4 neighbours.
The flag circuit in Figure~\ref{fig:checkz_2a2} can be performed on both processor topologies.
However, a full round of error syndrome extraction requires all the stabiliser generators of the Steane code to be measured.
The full syndrome extraction using only these two flag circuits (Figure~\ref{fig:checkz_2a1} and Figure~\ref{fig:checkz_2a2}) can be directly performed on the IBM-20 topology (e.g., a mapping in Figure~\ref{fig:steane2ibm222}) but not on the Surface-17 topology.

\begin{figure}[tbh!]
 \centering
     \begin{subfigure}[h]{0.2\textwidth}
    \includegraphics[width=\textwidth]{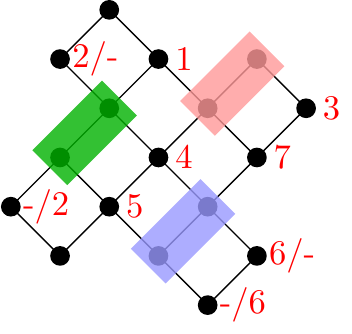}
    \caption{Steane-cl-L1}
    \label{fig:steane2sc17_222}
    \end{subfigure}
      \begin{subfigure}[h]{0.18\textwidth}
    \includegraphics[width=\textwidth]{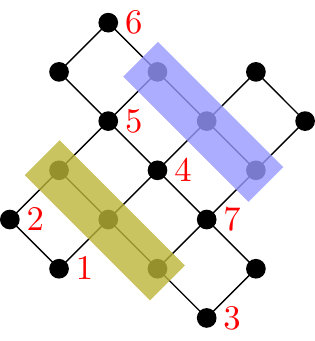}
    \caption{Steane-c2-L1}
    \label{fig:steane2sc17_33}
    \end{subfigure}
\caption{Mapping of the Steane code onto the Surface-17 topology, where the qubits labelled with numbers are data qubits and the qubits in the colored blocks are ancillas. (a) The mapping using the two-ancilla flag-bridge circuits in Figures~\ref{fig:checkz_2a2} and~\ref{fig:check4_2a_13} in which only one stabiliser is measured at a time; (b) The mapping using the three-ancilla flag-bridge circuits in Figures~\ref{fig:check4_3a_121} and~\ref{fig:check44_3a_222} that measure one and two stabilisers, respectively. 
}
\label{fig:steane2sc17}
\end{figure}

As mentioned above, all the flag-bridge circuits shown in this paper are used to measure $Z$ stabilisers, similar circuits with the same ancillas can be derived for measuring $X$-stabilisers.
Figures~\ref{fig:steane2sc17} and~\ref{fig:steane2ibm20} show examples of mapping the Steane code error correction using the flag-bridge circuits onto the Surface-17 topology and the IBM-20 topology, respectively.

In these mapping figures, the qubits in each red, blue, or green block are the ancillas in each flag-bridge circuit and they are used to measure the corresponding $Z$($X$)-stabiliser in the same color plaquette in Figure~\ref{fig:steaneqec}. 
The flag-bridge qubits in the yellow block are used to measure the $Z$($X$)-stabilisers in both red and green plaquettes.
The flag-bridge qubits in the gray block measure the $Z$($X$)-stabilisers in all three plaquettes.
The $X$ and $Z$ stabilisers are measured separately, more specifically, one first measures all the stabilisers in one type and then measures the other type. 
Furthermore, each of the flag-bridge circuits for the Steane code error correction need to be executed sequentially.
On the Surface-17 topology, one can measure all the stabilisers of the Steane code one by one when using the mapping in Figure~\ref{fig:steane2sc17_222}.
Maximally two stabilisers can be measured in parallel in this topology as shown in Figure~\ref{fig:steane2sc17_33}.
In contrast, three $Z$($X$)-stabilisers can be measured at the same time on the IBM-20 topology (Figure~\ref{fig:steane2ibm4}).

The circuit characterization of one full round of syndrome extraction for the Steane code when using different mappings is shown in Table~\ref{table:circuitcompare}.
This characterization includes the total number of ancilla qubits, the total number of operations and timesteps, and the number of f-\cnot and s-\cnot gates.
For comparison, we also show these parameters of the rotated distance-3 surface code (SC d=3).
As shown in Table~\ref{table:circuitcompare}, the circuits which can measure more stabilisers simultaneously require less operations and less timesteps.
Moreover, though the distance-3 surface code uses more ancilla qubits, it always needs less operations and less timesteps than the Steane code.


\begin{table}[tbh!]
\centering
\resizebox{0.48\textwidth}{!}{
\begin{tabular}{c|c|c|c|c|c}
\hline
              & \# Ancillas & \# Operations & \# f-CNOTs & \# s-CNOTs & \# Timesteps \\ \hline
Steane-c1-L1 & 6      & 72      & 12      & 24      & 50      \\ \hline
Steane-c1-L2 & 6      & 72      & 12      & 24      & 48      \\ \hline
Steane-c2-L1 & 6      & 72      & 16      & 20      & 40      \\ \hline
Steane-c2-L2 & 5      & 62      & 12      & 20      & 36      \\ \hline
Steane-c3-L2 & 4      & 54      & 12      & 18      & 26      \\ \hline
SC d=3       & 8      & 48      & 0      & 24      & 8      \\ \hline
\end{tabular}
}
\caption{Comparison of the quantum error correction circuits when different mappings are applied. 
}
\label{table:circuitcompare}
\end{table}

\begin{figure*}[tbh!]
 \centering
  \begin{subfigure}[h]{0.4\textwidth}
    \includegraphics[width=0.8\textwidth]{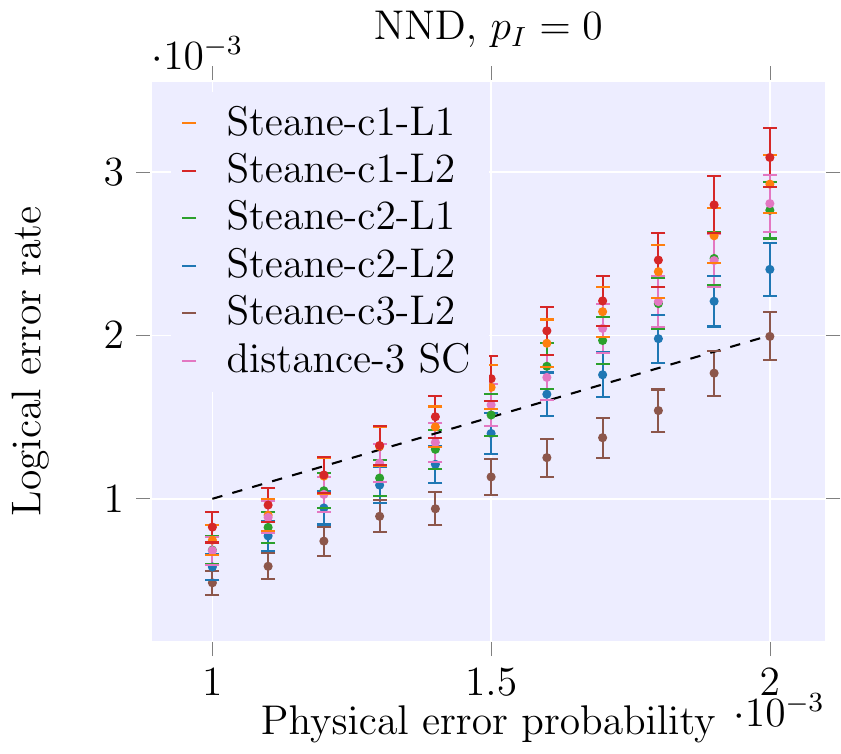}
    \caption{}
    \label{fig:lersteane_idle0}
  \end{subfigure}
  \begin{subfigure}[h]{0.4\textwidth}
   \includegraphics[width=0.8\textwidth]{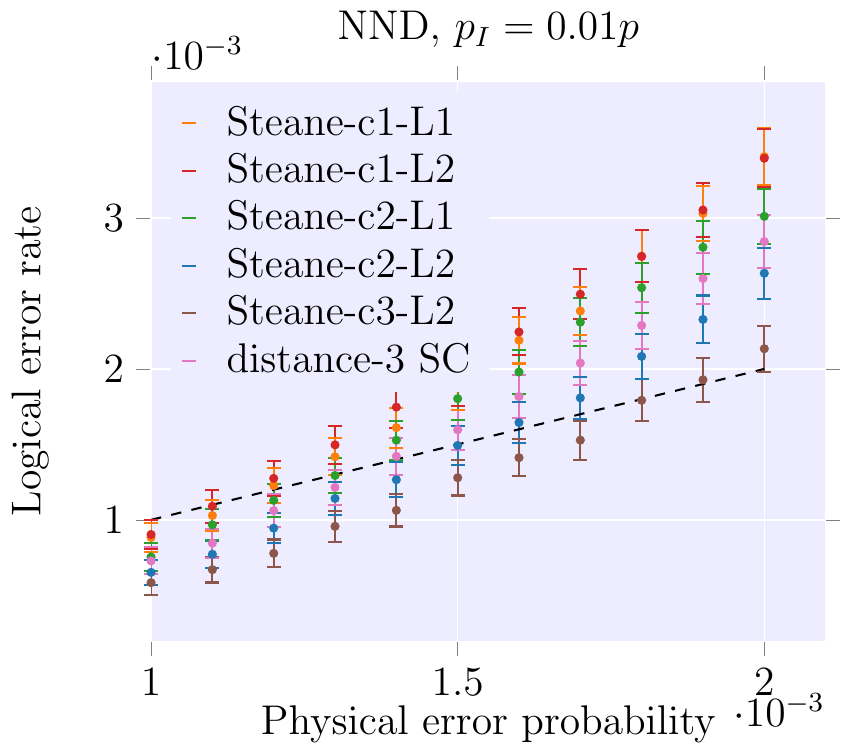}
   \caption{}
   \label{fig:lersteane_idle001}
  \end{subfigure}
  \begin{subfigure}[h]{0.4\textwidth}
   \includegraphics[width=0.8\textwidth]{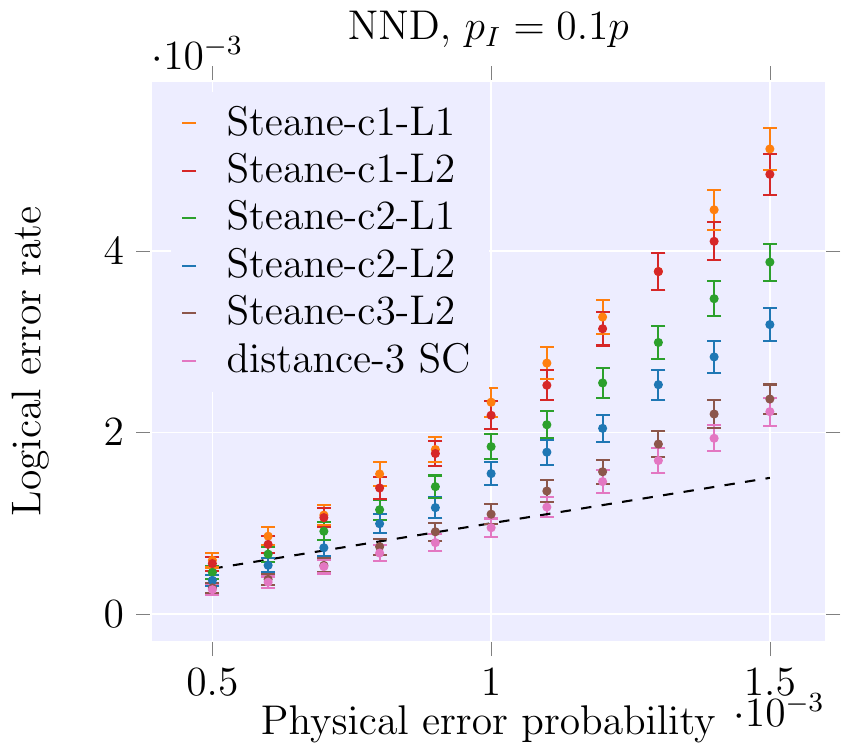}
   \caption{}
   \label{fig:lersteane_idle01}
  \end{subfigure}
  \begin{subfigure}[h]{0.4\textwidth}
   \includegraphics[width=0.8\textwidth]{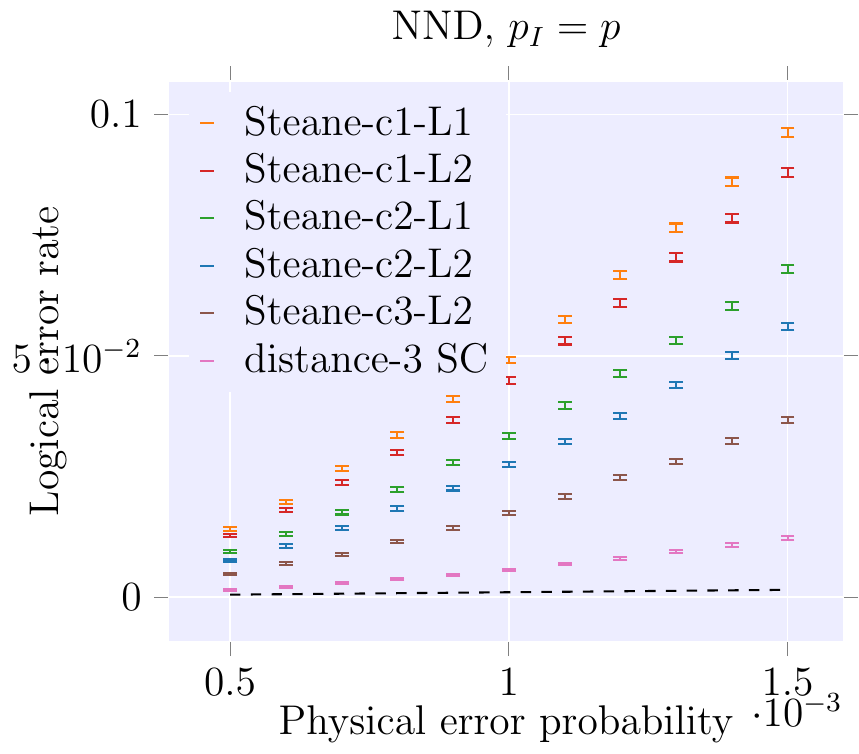}
   \caption{}
   \label{fig:lersteane_idle1}
  \end{subfigure}
\caption{Numerical simulation of the Steane code error correction based on different flag-bridge circuits using neural network decoders (NND). The circuit level noise ($p_{1}=p_{2}=p_{M}$) with ($p_{I} \neq 0$) or without idling errors ($p_{I} \neq 0$).}
\label{fig:lers_steane}
\end{figure*}\vspace{-5mm}


\subsection{Numerics}
We further compare different mapping circuits in terms of their fault tolerance, which is analyzed by numerical simulation under circuit level noise.
For each point in the numerics,
$10^{6}$ iterations of a full QEC cycle have been run and confidence intervals
at $99.9\%$ are plotted.
Moreover, NN decoders are used for this comparison since it has better performance than LUT decoders (see Figures~\ref{fig:lersteane_idle0} and ~\ref{fig:ler_lut}).
As shown in Figure~\ref{fig:lers_steane}, for the Steane code, the circuits that can measure more stabilisers in parallel have lower logical error rates, likely because they consist of fewer operations and require fewer timesteps.
Moreover, when there are no idling errors ($p_{I}=0$ in Figure~\ref{fig:lersteane_idle0}) or a small probability of idling errors ($p_{I}=0.01p$ in Figure~\ref{fig:lersteane_idle001}), the Steane code can achieve similar performance to, or even outperform, the distance-3 surface code by parallelizing stabiliser measurements.
This is because the circuit for the surface code error correction consists of more s-\cnot gates than the QEC circuits that can measure several stabilisers in parallel for the Steane code.
When idling errors are significant, we observe that the circuit with fewer timesteps results in lower logical error rates (as shown in Figures~\ref{fig:lersteane_idle01} and~\ref{fig:lersteane_idle1} for $p_{I}=0.1p$ and $p_{I}=p$ respectively).

\section{Other applications of the flag-bridge circuits}
\label{sec:applications}

In this section, we foresee some possible applications of the flag-bridge circuits including both fault-tolerant quantum error correction and fault-tolerant quantum computation.

\begin{figure}[tbh!]
 \centering
  \begin{subfigure}[h]{0.23\textwidth}
  \includegraphics[width=\textwidth]{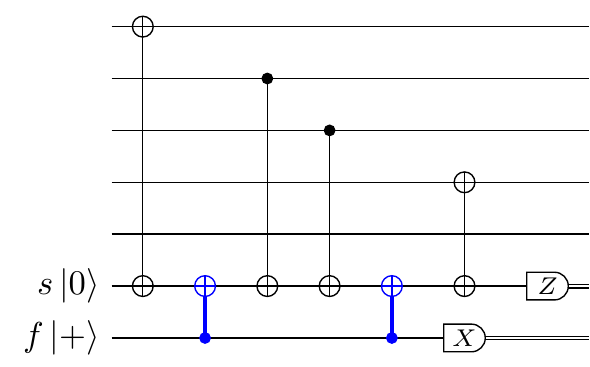}
  \caption{}
    \label{5code_2a_4}
    \end{subfigure}
  \begin{subfigure}[h]{0.23\textwidth}
     \includegraphics[width=\textwidth]{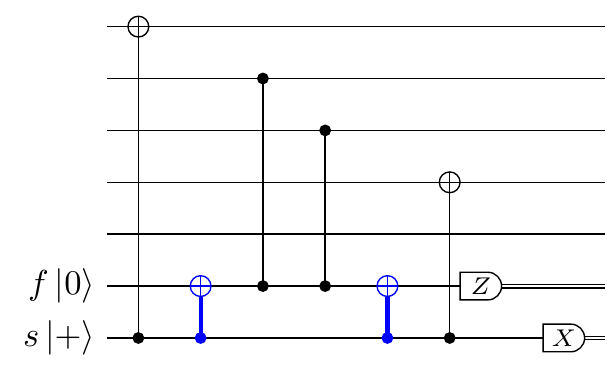}
     \caption{}
    \label{5code_2a_22_1}
    \end{subfigure}
  \begin{subfigure}[h]{0.23\textwidth}
     \includegraphics[width=\textwidth]{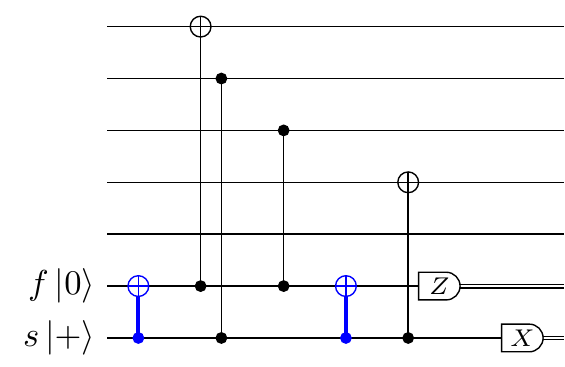}
     \caption{}
    \label{5code_2a_22_2}
    \end{subfigure}
  \begin{subfigure}[h]{0.23\textwidth}
\includegraphics[width=\textwidth]{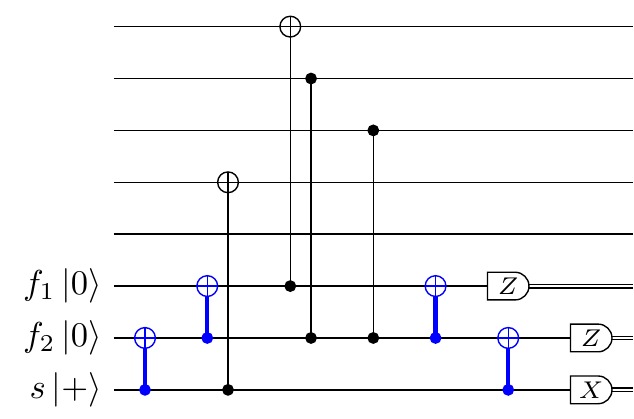}
\caption{}
    \label{5code_3a_121}
    \end{subfigure}
\caption{Fault-tolerant circuits for performing an $XZZX$-check: (a), (b), (c) using $2$ ancillas but requiring different connectivity; (d) using three ancillas, similar circuits can be generated by re-distributing the s-\cnot gates for each weight-4 check to different ancillas as mentioned in Section~\ref{sec:bridge}.}
\label{fig:5code_esm}
\end{figure}

\begin{figure}[tbh!]
 \centering
    \begin{subfigure}[b]{0.15\textwidth}
  \includegraphics[width=\textwidth]{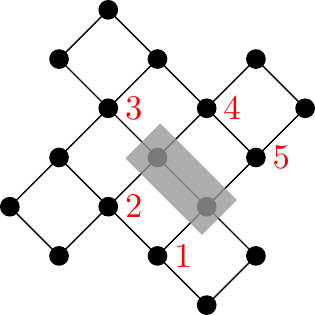}
  \caption{}
    \label{5code2sc17}
    \end{subfigure}
  \begin{subfigure}[b]{0.3\textwidth}
  \includegraphics[width=\textwidth]{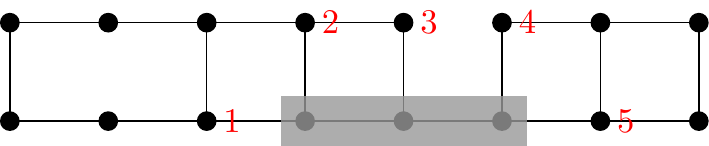}
  \caption{}
    \label{5code2ibm16}
    \end{subfigure}
\caption{Mapping of the 5-qubit code onto (a) the Surface-17 topology by using the two-ancilla flag-bridge circuits in Figure~\ref{fig:5code_esm} and (b) the IBM-16 topology using the three-ancilla circuit in Figure~\ref{5code_3a_121}.}
\label{fig:5code2device}
\end{figure}

\subsection{Flag-bridge QEC for the five-qubit code}
\label{sec:fivecode}
Analogous to the flag circuits, the flag-bridge circuits can also be applied to other distance-3 error correction codes such as the \qec{8}{3}{3}, \qec{10}{4}{3}, \qec{11}{5}{3}, \qec{5}{1}{3} codes, Hamming codes \qec{2^{r}-1}{2^{r}-1-2r}{3}, etc.
In this section, we consider the \qec{5}{1}{3} code as an example. This code has four stabilisers, which are cyclic permutations of $XZZXI$. 
Figure \ref{fig:5code_esm} shows the flag-bridge circuits that can measure an $XZZX$ stabiliser fault-tolerantly.
Each stabiliser of the 5-qubit code can be measured using these circuits up to data qubit permutation. 
Similar circuits using three ancillas to measure one stabiliser are also proposed in~\cite{yoder2017surface}.
All these circuits have different connectivity requirements.
By selecting and combining some of them, one can map the 5-qubit code error correction onto different qubit topologies.
Figure \ref{fig:5code2device} shows the mapping of the 5-qubit code to the Surface-17 processor topology using the two-ancilla flag-bridge circuits and the IBM Q Melbourne (IBM-16) processor topology using the three-ancilla flag-bridge circuits. 

\subsection{Flag-bridge circuits for FT computation}
\label{sec:computation}
\begin{figure}[tbh!]
\centering
\begin{subfigure}[b]{0.15\textwidth}
  \includegraphics[height=1.5\textwidth]{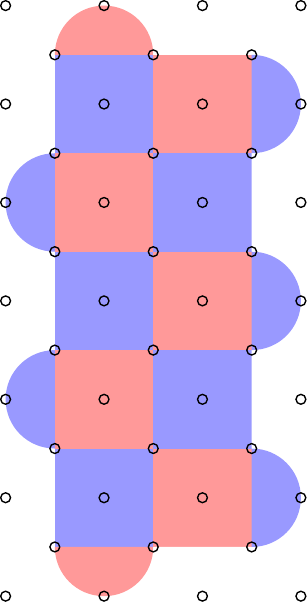}
   \caption{}
    \label{fig:sclayout_merge}
    \end{subfigure}
    \begin{subfigure}[b]{0.15\textwidth}
  \includegraphics[height=1.5\textwidth]{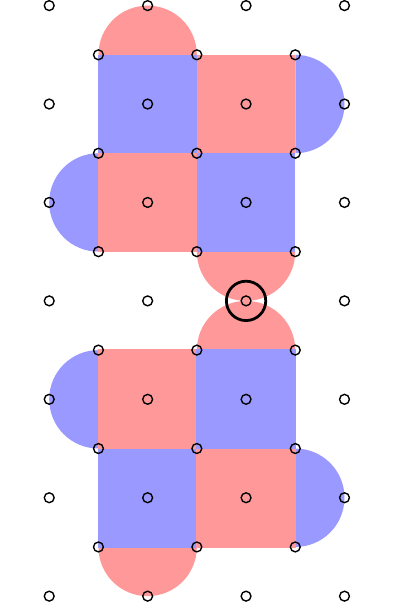}
   \caption{}
    \label{fig:sclayout_cnot1}
    \end{subfigure}
    \begin{subfigure}[b]{0.15\textwidth}
  \includegraphics[height=1.5\textwidth]{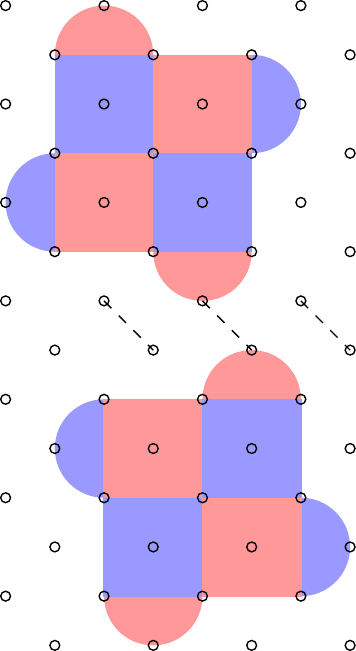}
   \caption{}
    \label{fig:sclayout_cnot2}
    \end{subfigure}
\caption{Mapping lattice surgery-based operations for the distance-3 surface code using flag-bridge circuits. Each red (blue) plaquette represents a weight-4 or weight-2 $X$($Z$)-stabiliser. Data qubits are on the vertices and ancilla qubits are on the plaquettes. (a) and (b) Initial layouts for performing a merge and a split operation using lattice surgery, respectively. (c) The layout after mapping using the two-ancilla flag-bridge circuits.}
\label{fig:sclayout_cnot}
\end{figure}

\begin{figure}[tbh!]
\centering
    \begin{subfigure}[b]{0.15\textwidth}
  \includegraphics[height=1.5\textwidth]{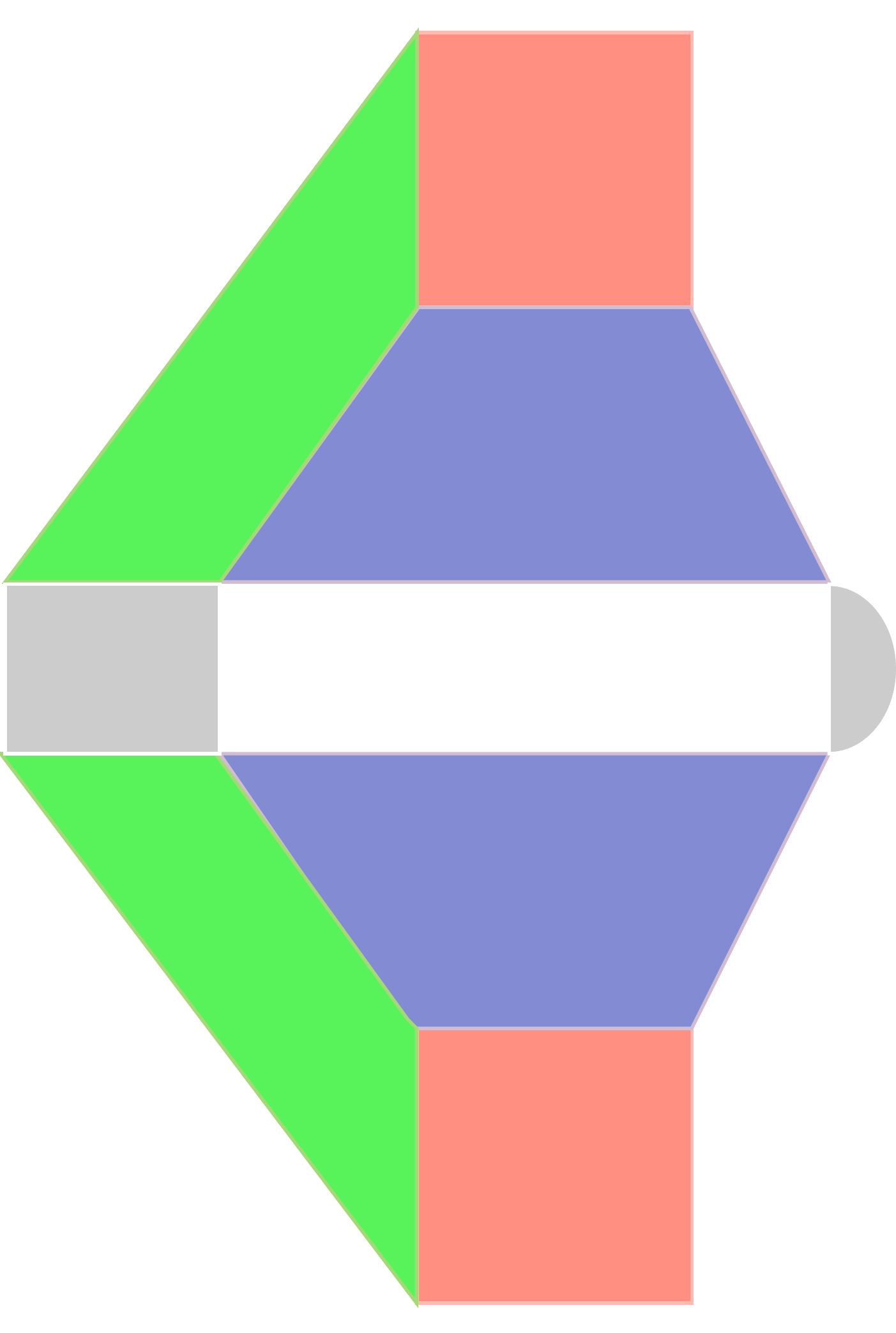}
   \caption{}
    \label{fig:steane_cnot1}
    \end{subfigure}\hspace{5mm}
    \begin{subfigure}[b]{0.15\textwidth}
  \includegraphics[height=1.5\textwidth]{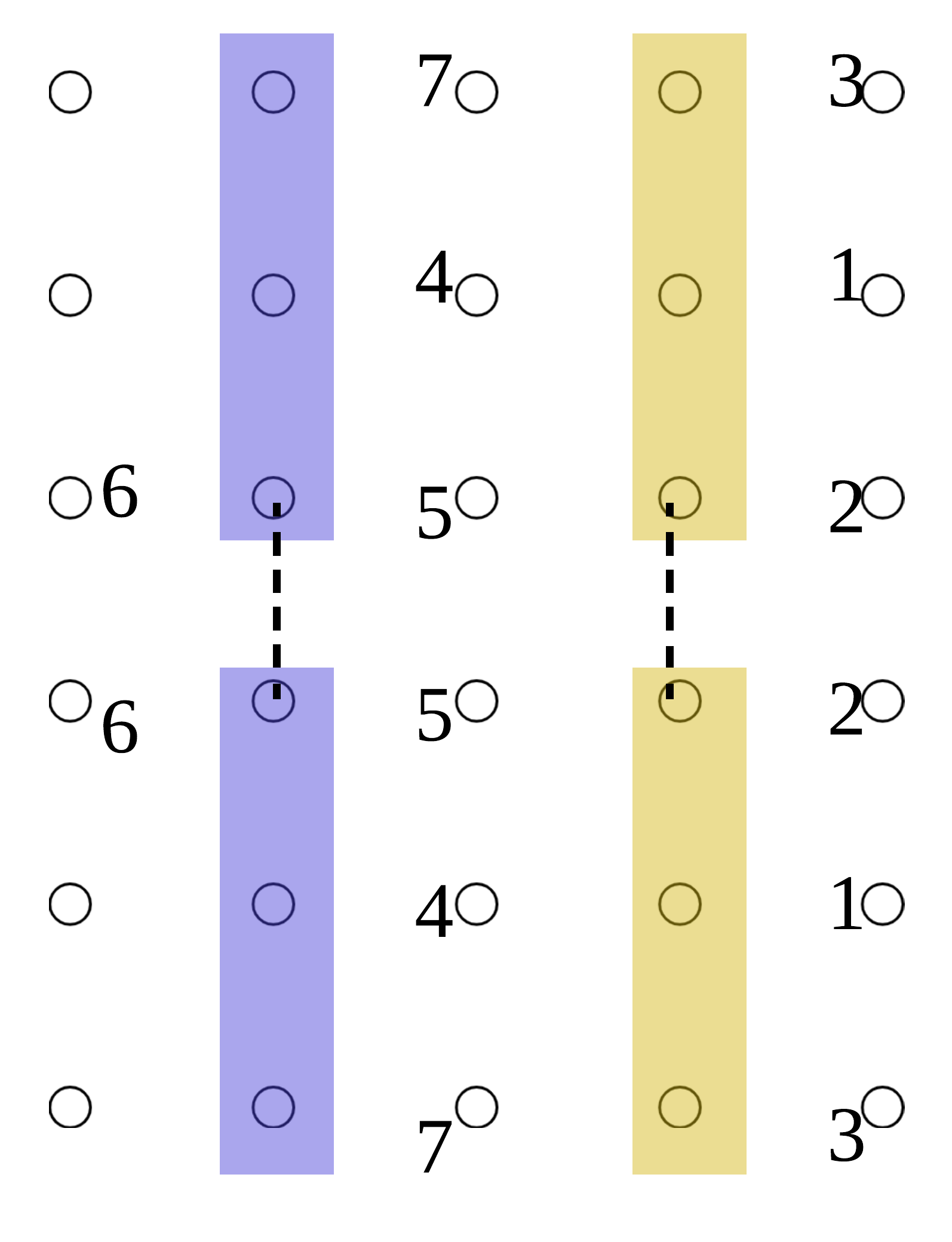}
   \caption{}
    \label{fig:steane_cnot2}
    \end{subfigure}
\caption{Mapping lattice surgery-based operations for the Steane code using flag-bridge circuits. (a) Initial layout, where 
the gray plaquettes only contain one type of stabilisers, depending on which joint measurement needs to be performed. Data qubits are on the vertices and ancilla qubits are on the plaquettes. (b) Mapping to a grid topology similar to Figure~\ref{fig:steane2sc17_33}. }
\label{fig:steane_cnot}
\end{figure}

The geometrical interaction constraint in near-term quantum processors has also limited the fault-tolerant implementation of logical operations.
For instance, a fault-tolerant \cnot gate in planar surface codes and color codes in principle can be implemented transversally in a 3D structure, that is, performing pair-wise \cnot gates between data qubits of the two lattices. 
However, this transversal \cnot is not realizable in near-term quantum technologies because of the local qubit connectivity limitation in a 2D architecture. 
Measurement-based protocols such as lattice surgery~\cite{horsman2012surface, landahl2014quantum} and code deformation~\cite{bombin2009quantum, vuillot2018code} have been proposed to comply with the 2D local interaction constraint.
Figures~\ref{fig:sclayout_cnot} and~\ref{fig:steane_cnot} show the qubit layouts for performing lattice-surgery-based operations on the distance-3 surface code and the distance-3 color code (the Steane code), respectively.
The details of implementing logical operations by lattice surgery can be found in~\cite{horsman2012surface, landahl2014quantum}.

As shown in Figure~\ref{fig:sclayout_cnot}, the merge operations can be directly performed on a 2D grid topology.
As mentioned previously, the stabiliser measurement of surface codes can be realized by only using the one-ancilla circuit similar to Figure~\ref{fig:checkz_1a}.
However, one ancilla qubit (the circled one in Figure~\ref{fig:sclayout_cnot1}) is used by two stabilisers from different lattices during the split operation.
One may have to measure these two stabilisers sequentially, which leads to more timesteps and in turn may result in higher logical error rates.
To preserve parallelism of the stabiliser measurement, we propose to use the qubit layout in Figure~\ref{fig:sclayout_cnot2}.
By using this layout, one can measure all the stabilisers in parallel when splitting lattices since they no longer share ancillas.
One can also perform the merge operation by replacing the original syndrome extraction circuit using one ancilla with the proposed flag-bridge circuits using two ancillas (Figure~\ref{fig:checkz_2a2}) where ancillas are connected by dash lines in Figure~\ref{fig:sclayout_cnot2} .
Similar mapping can be applied to other code-deformation-based operations on surface codes.

Furthermore, lattice-surgery-based operations for the Steane code in Figure~\ref{fig:steane_cnot1} cannot be directly realized in a 2D grid topology.
Similar to the mapping in Figure~\ref{fig:steane2sc17_33}, one can map these operations fault-tolerantly using the three-ancilla flag-bridge circuits as shown in Figure~\ref{fig:steane_cnot2}.
Compared to the distance-3 surface code, the Steane code can achieve Clifford gates transversally. 
Moreover, it requires fewer qubits for both FT error correction and FT computation, which may be preferable for demonstrating fault tolerance in small experiments.

\section{Discussion and conclusion}
\label{sec:conclude}

We have shown that the flag circuits can be phrased as one using encoded ancillas in an \qec{m}{p}{1} code. 
Based on this formulation, we proposed a flag-bridge approach to perform fault-tolerant quantum error correction for distance-3 codes on connectivity-constrained near-term quantum processors with low overhead.
Furthermore, we mapped the Steane code error correction onto two current qubit topologies using the flag-bridge circuits.
The numerical simulation results have shown that, 
the QEC circuits that can measure more stabilisers in parallel achieve lower logical error rates, providing insights for fabricating processors with more connectivity.
Moreover, we also showed that flag-bridge circuits can be applied to the 5-qubit code and lattice-surgery-based operations for the surface codes and the Steane code.
In addition, we have observed that the Steane code implementation that uses fewer qubits even outperforms the distance-3 surface code when idling errors occur with low probability.
The Steane code also allows transversal Clifford gates, which may make it a better candidate than the distance-3 surface code for demonstrating fault tolerance in small experiments.
However, the numerics in this work were carried out with Pauli errors, it will be interesting to test these circuits using more realistic error models.
Furthermore, the mapping procedure in this work was hand-optimized.
Future work will focus on automating the fault-tolerant mapping of flag-bridge quantum error correction onto given processors.
Besides, we also need to investigate the extensibility and scalability to higher distance codes and fault-tolerant computation.

\begin{acknowledgments}
The authors would like to thank Ben Criger for enlightening discussions on this project and feedback on the manuscript. 
We also thank Yang Wang and Xiaotong Ni for useful discussions on the implementation of error decoders. 
LLL acknowledges funding from the China Scholarship Council.
CGA acknowledges support from the Intel Corporation.
\end{acknowledgments}

\appendix
\section{Implementation of LUT and NN decoders}
\label{app:decoder}
Based on the FT QEC procedure for distance-3 codes in Section~\ref{sec:bridge}, decoding is only needed when two rounds of syndrome extraction (SE) are performed (the first round has non-trivial syndromes or flags).
If there is only non-trivial syndromes (no flags) in the first round, then the decoders will only decode using the measurement results in the second round. 
If there is any non-trivial flag in the first round, then the decoders will decode using these flags and the measurement results in the second round.
For the measurement information in the second round, the simple LUT decoder only considers the results of syndrome qubits, which is enough for correcting all the errors caused by one fault.
In contrast, the NN decoder also takes the flags of the second round into account.
This means the NN decoder could potentially correct some errors caused by more faults, outperforming the LUT decoder.

\begin{figure}[tbh!]
 \centering
    \includegraphics[width=0.35\textwidth]{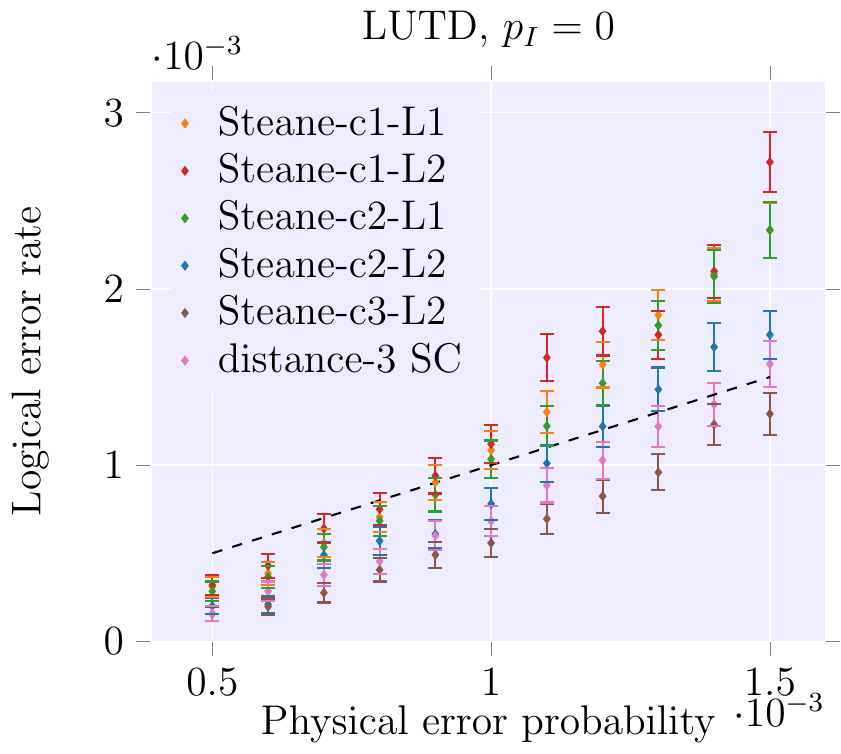}
\caption{Performance of the LUT decoder for the Steane code under circuit level noise without idling errors.}
\label{fig:ler_lut}
\end{figure}

\begin{table}[tbh!]
\centering
\resizebox{0.5\textwidth}{!}{
\begin{tabular}{|c|c|c|c|c|c|c|c|c|}
\hline
\multirow{2}{*}{\begin{tabular}[c]{@{}c@{}}Loss \\ function\end{tabular}} & \multirow{2}{*}{\begin{tabular}[c]{@{}c@{}}Hidden \\ layers\end{tabular}} & \multicolumn{2}{c|}{Activation function}                                                                          & \multirow{2}{*}{Optimizer}                             & \multirow{2}{*}{\begin{tabular}[c]{@{}c@{}}Learning \\ rate\end{tabular}} & \multirow{2}{*}{\begin{tabular}[c]{@{}c@{}}Batch \\ size\end{tabular}} & \multirow{2}{*}{PER} & \multirow{2}{*}{Samples} \\ \cline{3-4}
                                                                          &                                                                              & \begin{tabular}[c]{@{}c@{}}Output\\  layer\end{tabular} & \begin{tabular}[c]{@{}c@{}}hidden \\ layer\end{tabular} &                                                        &                                                                           &                                                                        &                      &                             \\ \hline
\begin{tabular}[c]{@{}c@{}}cross-\\ entropy\end{tabular}                  & 3                                                                            & sigmoid                                                 & \begin{tabular}[c]{@{}c@{}}ReLU\\  (tanh)\end{tabular}  & \begin{tabular}[c]{@{}c@{}}Adam\\ (Nadam)\end{tabular} & 0.002                                                                     & 50                                                                     & $\sim 0.01$          & $10^{5}$ \\ \hline
\end{tabular}
}
\caption{The implementation details of the NN decoder.}
\label{tbl:nn}
\end{table}


\textbf{The LUT decoder:}
As mentioned previously, we use a brute-force search to check the fault tolerance of flag-bridge circuits.
After this search, all the errors from one single fault and the corresponding syndrome-flag (SF) string are collected.
For FT flag-bridge circuits, these error-SF pairs can be directly used to design a LUT decoder.
Two look-up tables need to be created. One is used for the case where only syndromes are observed in the first round of SE with a size $2^{m_{s}}$, $m_{s}$ is the total number of syndrome qubits in the QEC circuit  $\mathcal{C} =\left \{ c(g_{1}), c(g_{2}),\cdots, c(g_{r}) \right \}$.
Note that if the same ancilla qubits are re-used in different $c_{g}$, they are still considered as different syndrome qubits, similarly for flag qubits. 
The other table is to decode for the case where flags are raised in the first round of SE, which has a size of
$\sum_{i}2^{m_{f_{i}}}2^{m_{s}}$, $m_{f_{i}}$ is the total number of flag qubits in $c_{g_{i}}$.
The LUT decoder is designed to correct all single faults, but not to correct the most likely two faults correspond to measured syndromes.
The performance of different flag-bridge circuits for the Steane code using LUT decoders is shown in Figure~\ref{fig:ler_lut}.
As can be seen, the QEC circuits that can achieve more parallelism of stabiliser measurement have lower logical error rates.

\textbf{The NN decoder:}
Decoding can be seen as a classification problem, that is, given the observed syndromes, the decoder identifies the error or the logical coset of the error that has occurred .
It has been shown that neural networks are versatile
tools for decoding topological quantum error correction codes \cite{krastanov2017deep, varsamopoulos2017decoding, baireuther2018machine, ni2018neural}.
The inputs $x_{i}$ for a neural network decoder are the syndromes (and flags for flag QEC).
In this paper, two rounds of syndromes and flags will be collected when using the flag-bridge error correction for distance-3 codes.
Therefore, the size of input layer will be $2\times m$, $m$ is the total number of syndrome and flag-bridge qubits.
In this work, the outputs $y_{i}$ are the suggested physical errors which can result in the given syndromes and flags. 
For a CSS code with $n$ data qubits, the size of output layer is set to be $2\times n$, which can describe whether a $X$ or/and a $Z$ error has occurred on each data qubit.
The neural network will find an approximate function $f: x \rightarrow y$ to describe the input-output relation from the set of training data $\{(x_{i}, y_{i})\}$.
Note for large-distance codes, it is more efficient to use logical errors as outputs and a simple decoder (e.g., LUT decoder) is required to generate the logical error information.

In this work, a simple NN decoder using the Tensorflow library \cite{abadi2016tensorflow} is developed to analyze the fault tolerance of different flag-bridge circuits. 
We use the `sigmoid' activation function for the output layer and $10^{5}$ syndrome-error pairs at physical error rate (PER) around $0.01$ are sampled for each training, more details of the designed NN decoder are described in Table~\ref{tbl:nn}.
Since the focus of this work is to evaluate the flag-bridge quantum error correction, we leave the performance and speed optimization of NN decoders for future work.

\bibliography{references}

\begin{thebibliography}{40}%
\makeatletter
\providecommand \@ifxundefined [1]{%
 \@ifx{#1\undefined}
}%
\providecommand \@ifnum [1]{%
 \ifnum #1\expandafter \@firstoftwo
 \else \expandafter \@secondoftwo
 \fi
}%
\providecommand \@ifx [1]{%
 \ifx #1\expandafter \@firstoftwo
 \else \expandafter \@secondoftwo
 \fi
}%
\providecommand \natexlab [1]{#1}%
\providecommand \enquote  [1]{``#1''}%
\providecommand \bibnamefont  [1]{#1}%
\providecommand \bibfnamefont [1]{#1}%
\providecommand \citenamefont [1]{#1}%
\providecommand \href@noop [0]{\@secondoftwo}%
\providecommand \href [0]{\begingroup \@sanitize@url \@href}%
\providecommand \@href[1]{\@@startlink{#1}\@@href}%
\providecommand \@@href[1]{\endgroup#1\@@endlink}%
\providecommand \@sanitize@url [0]{\catcode `\\12\catcode `\$12\catcode
  `\&12\catcode `\#12\catcode `\^12\catcode `\_12\catcode `\%12\relax}%
\providecommand \@@startlink[1]{}%
\providecommand \@@endlink[0]{}%
\providecommand \url  [0]{\begingroup\@sanitize@url \@url }%
\providecommand \@url [1]{\endgroup\@href {#1}{\urlprefix }}%
\providecommand \urlprefix  [0]{URL }%
\providecommand \Eprint [0]{\href }%
\providecommand \doibase [0]{http://dx.doi.org/}%
\providecommand \selectlanguage [0]{\@gobble}%
\providecommand \bibinfo  [0]{\@secondoftwo}%
\providecommand \bibfield  [0]{\@secondoftwo}%
\providecommand \translation [1]{[#1]}%
\providecommand \BibitemOpen [0]{}%
\providecommand \bibitemStop [0]{}%
\providecommand \bibitemNoStop [0]{.\EOS\space}%
\providecommand \EOS [0]{\spacefactor3000\relax}%
\providecommand \BibitemShut  [1]{\csname bibitem#1\endcsname}%
\let\auto@bib@innerbib\@empty
\bibitem [{\citenamefont {Preskill}(2018)}]{preskill2018quantum}%
  \BibitemOpen
  \bibfield  {author} {\bibinfo {author} {\bibfnamefont {J.}~\bibnamefont
  {Preskill}},\ }\href@noop {} {\bibfield  {journal} {\bibinfo  {journal}
  {Quantum}\ }\textbf {\bibinfo {volume} {2}},\ \bibinfo {pages} {79} (\bibinfo
  {year} {2018})}\BibitemShut {NoStop}%
\bibitem [{\citenamefont {Kelly}(2018)}]{google72qubit}%
  \BibitemOpen
  \bibfield  {author} {\bibinfo {author} {\bibfnamefont {J.}~\bibnamefont
  {Kelly}},\ }\href@noop {} {\emph {\bibinfo {title} {A Preview of
  {Bristlecone}, Google’s New Quantum Processor}}},\ \bibinfo {type} {News
  from Google AI}\ (\bibinfo  {institution} {Google LLC},\ \bibinfo {year}
  {2018})\BibitemShut {NoStop}%
\bibitem [{\citenamefont {Knight}(2018)}]{ibm50qubit}%
  \BibitemOpen
  \bibfield  {author} {\bibinfo {author} {\bibfnamefont {W.}~\bibnamefont
  {Knight}},\ }\href@noop {} {\emph {\bibinfo {title} {{IBM} Raises the Bar
  with a 50-Qubit Quantum Computer}}},\ \bibinfo {type} {News}\ (\bibinfo
  {institution} {MIT Technology Review},\ \bibinfo {year} {2018})\BibitemShut
  {NoStop}%
\bibitem [{\citenamefont {Hsu}(2018)}]{intel49qubit}%
  \BibitemOpen
  \bibfield  {author} {\bibinfo {author} {\bibfnamefont {J.}~\bibnamefont
  {Hsu}},\ }\href@noop {} {\emph {\bibinfo {title} {{CES 2018}: Intel's
  49-Qubit Chip Shoots for Quantum Supremacy}}},\ \bibinfo {type} {General
  technology blog}\ (\bibinfo  {institution} {IEEE Spectrum},\ \bibinfo {year}
  {2018})\BibitemShut {NoStop}%
\bibitem [{\citenamefont {Barends}\ \emph {et~al.}(2014)\citenamefont
  {Barends}, \citenamefont {Kelly}, \citenamefont {Megrant}, \citenamefont
  {Veitia}, \citenamefont {Sank}, \citenamefont {Jeffrey}, \citenamefont
  {White}, \citenamefont {Mutus}, \citenamefont {Fowler}, \citenamefont
  {Campbell} \emph {et~al.}}]{barends2014superconducting}%
  \BibitemOpen
  \bibfield  {author} {\bibinfo {author} {\bibfnamefont {R.}~\bibnamefont
  {Barends}}, \bibinfo {author} {\bibfnamefont {J.}~\bibnamefont {Kelly}},
  \bibinfo {author} {\bibfnamefont {A.}~\bibnamefont {Megrant}}, \bibinfo
  {author} {\bibfnamefont {A.}~\bibnamefont {Veitia}}, \bibinfo {author}
  {\bibfnamefont {D.}~\bibnamefont {Sank}}, \bibinfo {author} {\bibfnamefont
  {E.}~\bibnamefont {Jeffrey}}, \bibinfo {author} {\bibfnamefont {T.~C.}\
  \bibnamefont {White}}, \bibinfo {author} {\bibfnamefont {J.}~\bibnamefont
  {Mutus}}, \bibinfo {author} {\bibfnamefont {A.~G.}\ \bibnamefont {Fowler}},
  \bibinfo {author} {\bibfnamefont {B.}~\bibnamefont {Campbell}},  \emph
  {et~al.},\ }\href@noop {} {\bibfield  {journal} {\bibinfo  {journal}
  {Nature}\ }\textbf {\bibinfo {volume} {508}},\ \bibinfo {pages} {500}
  (\bibinfo {year} {2014})}\BibitemShut {NoStop}%
\bibitem [{\citenamefont {Hempel}\ \emph {et~al.}(2018)\citenamefont {Hempel},
  \citenamefont {Maier}, \citenamefont {Romero}, \citenamefont {McClean},
  \citenamefont {Monz}, \citenamefont {Shen}, \citenamefont {Jurcevic},
  \citenamefont {Lanyon}, \citenamefont {Love}, \citenamefont {Babbush} \emph
  {et~al.}}]{hempel2018quantum}%
  \BibitemOpen
  \bibfield  {author} {\bibinfo {author} {\bibfnamefont {C.}~\bibnamefont
  {Hempel}}, \bibinfo {author} {\bibfnamefont {C.}~\bibnamefont {Maier}},
  \bibinfo {author} {\bibfnamefont {J.}~\bibnamefont {Romero}}, \bibinfo
  {author} {\bibfnamefont {J.}~\bibnamefont {McClean}}, \bibinfo {author}
  {\bibfnamefont {T.}~\bibnamefont {Monz}}, \bibinfo {author} {\bibfnamefont
  {H.}~\bibnamefont {Shen}}, \bibinfo {author} {\bibfnamefont {P.}~\bibnamefont
  {Jurcevic}}, \bibinfo {author} {\bibfnamefont {B.~P.}\ \bibnamefont
  {Lanyon}}, \bibinfo {author} {\bibfnamefont {P.}~\bibnamefont {Love}},
  \bibinfo {author} {\bibfnamefont {R.}~\bibnamefont {Babbush}},  \emph
  {et~al.},\ }\href@noop {} {\bibfield  {journal} {\bibinfo  {journal} {Phys.
  Rev. X}\ }\textbf {\bibinfo {volume} {8}},\ \bibinfo {pages} {031022}
  (\bibinfo {year} {2018})}\BibitemShut {NoStop}%
\bibitem [{\citenamefont {Kokail}\ \emph {et~al.}(2019)\citenamefont {Kokail},
  \citenamefont {Maier}, \citenamefont {van Bijnen}, \citenamefont {Brydges},
  \citenamefont {Joshi}, \citenamefont {Jurcevic}, \citenamefont {Muschik},
  \citenamefont {Silvi}, \citenamefont {Blatt}, \citenamefont {Roos} \emph
  {et~al.}}]{kokail2019self}%
  \BibitemOpen
  \bibfield  {author} {\bibinfo {author} {\bibfnamefont {C.}~\bibnamefont
  {Kokail}}, \bibinfo {author} {\bibfnamefont {C.}~\bibnamefont {Maier}},
  \bibinfo {author} {\bibfnamefont {R.}~\bibnamefont {van Bijnen}}, \bibinfo
  {author} {\bibfnamefont {T.}~\bibnamefont {Brydges}}, \bibinfo {author}
  {\bibfnamefont {M.}~\bibnamefont {Joshi}}, \bibinfo {author} {\bibfnamefont
  {P.}~\bibnamefont {Jurcevic}}, \bibinfo {author} {\bibfnamefont
  {C.}~\bibnamefont {Muschik}}, \bibinfo {author} {\bibfnamefont
  {P.}~\bibnamefont {Silvi}}, \bibinfo {author} {\bibfnamefont
  {R.}~\bibnamefont {Blatt}}, \bibinfo {author} {\bibfnamefont
  {C.}~\bibnamefont {Roos}},  \emph {et~al.},\ }\href@noop {} {\bibfield
  {journal} {\bibinfo  {journal} {Nature}\ }\textbf {\bibinfo {volume} {569}},\
  \bibinfo {pages} {355} (\bibinfo {year} {2019})}\BibitemShut {NoStop}%
\bibitem [{\citenamefont {Fu}\ \emph {et~al.}(2016)\citenamefont {Fu},
  \citenamefont {Riesebos}, \citenamefont {Lao}, \citenamefont {Almudever},
  \citenamefont {Sebastiano}, \citenamefont {Versluis}, \citenamefont
  {Charbon},\ and\ \citenamefont {Bertels}}]{fu2016heterogeneous}%
  \BibitemOpen
  \bibfield  {author} {\bibinfo {author} {\bibfnamefont {X.}~\bibnamefont
  {Fu}}, \bibinfo {author} {\bibfnamefont {L.}~\bibnamefont {Riesebos}},
  \bibinfo {author} {\bibfnamefont {L.}~\bibnamefont {Lao}}, \bibinfo {author}
  {\bibfnamefont {C.}~\bibnamefont {Almudever}}, \bibinfo {author}
  {\bibfnamefont {F.}~\bibnamefont {Sebastiano}}, \bibinfo {author}
  {\bibfnamefont {R.}~\bibnamefont {Versluis}}, \bibinfo {author}
  {\bibfnamefont {E.}~\bibnamefont {Charbon}}, \ and\ \bibinfo {author}
  {\bibfnamefont {K.}~\bibnamefont {Bertels}},\ }in\ \href@noop {} {\emph
  {\bibinfo {booktitle} {Proceedings of the ACM International Conference on
  Computing Frontiers (CF)}}}\ (\bibinfo {organization} {ACM},\ \bibinfo {year}
  {2016})\ pp.\ \bibinfo {pages} {323--330}\BibitemShut {NoStop}%
\bibitem [{\citenamefont {Fu}\ \emph {et~al.}(2017)\citenamefont {Fu},
  \citenamefont {Rol}, \citenamefont {Bultink}, \citenamefont {Van~Someren},
  \citenamefont {Khammassi}, \citenamefont {Ashraf}, \citenamefont {Vermeulen},
  \citenamefont {De~Sterke}, \citenamefont {Vlothuizen}, \citenamefont
  {Schouten} \emph {et~al.}}]{fu2017experimental}%
  \BibitemOpen
  \bibfield  {author} {\bibinfo {author} {\bibfnamefont {X.}~\bibnamefont
  {Fu}}, \bibinfo {author} {\bibfnamefont {M.}~\bibnamefont {Rol}}, \bibinfo
  {author} {\bibfnamefont {C.}~\bibnamefont {Bultink}}, \bibinfo {author}
  {\bibfnamefont {J.}~\bibnamefont {Van~Someren}}, \bibinfo {author}
  {\bibfnamefont {N.}~\bibnamefont {Khammassi}}, \bibinfo {author}
  {\bibfnamefont {I.}~\bibnamefont {Ashraf}}, \bibinfo {author} {\bibfnamefont
  {R.}~\bibnamefont {Vermeulen}}, \bibinfo {author} {\bibfnamefont
  {J.}~\bibnamefont {De~Sterke}}, \bibinfo {author} {\bibfnamefont
  {W.}~\bibnamefont {Vlothuizen}}, \bibinfo {author} {\bibfnamefont
  {R.}~\bibnamefont {Schouten}},  \emph {et~al.},\ }in\ \href@noop {} {\emph
  {\bibinfo {booktitle} {Proceedings of the 50th Annual IEEE/ACM International
  Symposium on Microarchitecture}}}\ (\bibinfo {organization} {ACM},\ \bibinfo
  {year} {2017})\ pp.\ \bibinfo {pages} {813--825}\BibitemShut {NoStop}%
\bibitem [{\citenamefont {Shor}(1996)}]{shor1996fault}%
  \BibitemOpen
  \bibfield  {author} {\bibinfo {author} {\bibfnamefont {P.~W.}\ \bibnamefont
  {Shor}},\ }in\ \href@noop {} {\emph {\bibinfo {booktitle} {Proceedings of
  37th Conference on Foundations of Computer Science}}}\ (\bibinfo
  {organization} {IEEE},\ \bibinfo {year} {1996})\ pp.\ \bibinfo {pages}
  {56--65}\BibitemShut {NoStop}%
\bibitem [{\citenamefont {Steane}(1997)}]{steane1997active}%
  \BibitemOpen
  \bibfield  {author} {\bibinfo {author} {\bibfnamefont {A.~M.}\ \bibnamefont
  {Steane}},\ }\href@noop {} {\bibfield  {journal} {\bibinfo  {journal} {Phys.
  Rev. Lett.}\ }\textbf {\bibinfo {volume} {78}},\ \bibinfo {pages} {2252}
  (\bibinfo {year} {1997})}\BibitemShut {NoStop}%
\bibitem [{\citenamefont {Knill}(2005)}]{knill2005scalable}%
  \BibitemOpen
  \bibfield  {author} {\bibinfo {author} {\bibfnamefont {E.}~\bibnamefont
  {Knill}},\ }\href@noop {} {\bibfield  {journal} {\bibinfo  {journal} {Phys.
  Rev. A}\ }\textbf {\bibinfo {volume} {71}},\ \bibinfo {pages} {042322}
  (\bibinfo {year} {2005})}\BibitemShut {NoStop}%
\bibitem [{\citenamefont {Yoder}\ and\ \citenamefont
  {Kim}(2017)}]{yoder2017surface}%
  \BibitemOpen
  \bibfield  {author} {\bibinfo {author} {\bibfnamefont {T.~J.}\ \bibnamefont
  {Yoder}}\ and\ \bibinfo {author} {\bibfnamefont {I.~H.}\ \bibnamefont
  {Kim}},\ }\href@noop {} {\bibfield  {journal} {\bibinfo  {journal} {Quantum}\
  }\textbf {\bibinfo {volume} {1}},\ \bibinfo {pages} {2} (\bibinfo {year}
  {2017})}\BibitemShut {NoStop}%
\bibitem [{\citenamefont {Chao}\ and\ \citenamefont
  {Reichardt}(2018)}]{chao2018quantum}%
  \BibitemOpen
  \bibfield  {author} {\bibinfo {author} {\bibfnamefont {R.}~\bibnamefont
  {Chao}}\ and\ \bibinfo {author} {\bibfnamefont {B.~W.}\ \bibnamefont
  {Reichardt}},\ }\href@noop {} {\bibfield  {journal} {\bibinfo  {journal}
  {Phys. Rev. Lett.}\ }\textbf {\bibinfo {volume} {121}},\ \bibinfo {pages}
  {050502} (\bibinfo {year} {2018})}\BibitemShut {NoStop}%
\bibitem [{\citenamefont {Chamberland}\ and\ \citenamefont
  {Beverland}(2018)}]{chamberland2018flag}%
  \BibitemOpen
  \bibfield  {author} {\bibinfo {author} {\bibfnamefont {C.}~\bibnamefont
  {Chamberland}}\ and\ \bibinfo {author} {\bibfnamefont {M.~E.}\ \bibnamefont
  {Beverland}},\ }\href {\doibase 10.22331/q-2018-02-08-53} {\bibfield
  {journal} {\bibinfo  {journal} {Quantum}\ }\textbf {\bibinfo {volume} {2}},\
  \bibinfo {pages} {53} (\bibinfo {year} {2018})}\BibitemShut {NoStop}%
\bibitem [{\citenamefont {Reichardt}(2018)}]{reichardt2018fault}%
  \BibitemOpen
  \bibfield  {author} {\bibinfo {author} {\bibfnamefont {B.~W.}\ \bibnamefont
  {Reichardt}},\ }\href@noop {} {\bibfield  {journal} {\bibinfo  {journal}
  {arXiv:1804.06995}\ } (\bibinfo {year} {2018})}\BibitemShut {NoStop}%
\bibitem [{\citenamefont {IBM}(2017)}]{ibm17experience}%
  \BibitemOpen
  \bibfield  {author} {\bibinfo {author} {\bibnamefont {IBM}},\ }\href
  {https://www.research.ibm.com/ibm-q/} {\enquote {\bibinfo {title} {Quantum
  experience},}\ } (\bibinfo {year} {2017})\BibitemShut {NoStop}%
\bibitem [{\citenamefont {Versluis}\ \emph {et~al.}(2017)\citenamefont
  {Versluis}, \citenamefont {Poletto}, \citenamefont {Khammassi}, \citenamefont
  {Tarasinski}, \citenamefont {Haider}, \citenamefont {Michalak}, \citenamefont
  {Bruno}, \citenamefont {Bertels},\ and\ \citenamefont
  {DiCarlo}}]{versluis2017scalable}%
  \BibitemOpen
  \bibfield  {author} {\bibinfo {author} {\bibfnamefont {R.}~\bibnamefont
  {Versluis}}, \bibinfo {author} {\bibfnamefont {S.}~\bibnamefont {Poletto}},
  \bibinfo {author} {\bibfnamefont {N.}~\bibnamefont {Khammassi}}, \bibinfo
  {author} {\bibfnamefont {B.}~\bibnamefont {Tarasinski}}, \bibinfo {author}
  {\bibfnamefont {N.}~\bibnamefont {Haider}}, \bibinfo {author} {\bibfnamefont
  {D.}~\bibnamefont {Michalak}}, \bibinfo {author} {\bibfnamefont
  {A.}~\bibnamefont {Bruno}}, \bibinfo {author} {\bibfnamefont
  {K.}~\bibnamefont {Bertels}}, \ and\ \bibinfo {author} {\bibfnamefont
  {L.}~\bibnamefont {DiCarlo}},\ }\href@noop {} {\bibfield  {journal} {\bibinfo
   {journal} {Phys. Rev. Applied}\ }\textbf {\bibinfo {volume} {8}},\ \bibinfo
  {pages} {034021} (\bibinfo {year} {2017})}\BibitemShut {NoStop}%
\bibitem [{\citenamefont {Rigetti}(2018)}]{rigetti}%
  \BibitemOpen
  \bibfield  {author} {\bibinfo {author} {\bibnamefont {Rigetti}},\ }\href
  {https://www.rigetti.com/forest} {\enquote {\bibinfo {title} {Regetti
  forest},}\ } (\bibinfo {year} {2018})\BibitemShut {NoStop}%
\bibitem [{\citenamefont {DiVincenzo}\ and\ \citenamefont
  {Aliferis}(2007)}]{divincenzo2007effective}%
  \BibitemOpen
  \bibfield  {author} {\bibinfo {author} {\bibfnamefont {D.~P.}\ \bibnamefont
  {DiVincenzo}}\ and\ \bibinfo {author} {\bibfnamefont {P.}~\bibnamefont
  {Aliferis}},\ }\href@noop {} {\bibfield  {journal} {\bibinfo  {journal}
  {Phys. Rev. Lett.}\ }\textbf {\bibinfo {volume} {98}},\ \bibinfo {pages}
  {020501} (\bibinfo {year} {2007})}\BibitemShut {NoStop}%
\bibitem [{\citenamefont {Fowler}\ \emph {et~al.}(2012)\citenamefont {Fowler},
  \citenamefont {Mariantoni}, \citenamefont {Martinis},\ and\ \citenamefont
  {Cleland}}]{fowler2012surface}%
  \BibitemOpen
  \bibfield  {author} {\bibinfo {author} {\bibfnamefont {A.~G.}\ \bibnamefont
  {Fowler}}, \bibinfo {author} {\bibfnamefont {M.}~\bibnamefont {Mariantoni}},
  \bibinfo {author} {\bibfnamefont {J.~M.}\ \bibnamefont {Martinis}}, \ and\
  \bibinfo {author} {\bibfnamefont {A.~N.}\ \bibnamefont {Cleland}},\
  }\href@noop {} {\bibfield  {journal} {\bibinfo  {journal} {Phys. Rev. A}\
  }\textbf {\bibinfo {volume} {86}},\ \bibinfo {pages} {032324} (\bibinfo
  {year} {2012})}\BibitemShut {NoStop}%
\bibitem [{\citenamefont {Riste}\ \emph {et~al.}(2015)\citenamefont {Riste},
  \citenamefont {Poletto}, \citenamefont {Huang}, \citenamefont {Bruno},
  \citenamefont {Vesterinen}, \citenamefont {Saira},\ and\ \citenamefont
  {DiCarlo}}]{riste2015detecting}%
  \BibitemOpen
  \bibfield  {author} {\bibinfo {author} {\bibfnamefont {D.}~\bibnamefont
  {Riste}}, \bibinfo {author} {\bibfnamefont {S.}~\bibnamefont {Poletto}},
  \bibinfo {author} {\bibfnamefont {M.-Z.}\ \bibnamefont {Huang}}, \bibinfo
  {author} {\bibfnamefont {A.}~\bibnamefont {Bruno}}, \bibinfo {author}
  {\bibfnamefont {V.}~\bibnamefont {Vesterinen}}, \bibinfo {author}
  {\bibfnamefont {O.-P.}\ \bibnamefont {Saira}}, \ and\ \bibinfo {author}
  {\bibfnamefont {L.}~\bibnamefont {DiCarlo}},\ }\href@noop {} {\bibfield
  {journal} {\bibinfo  {journal} {Nature communications}\ }\textbf {\bibinfo
  {volume} {6}},\ \bibinfo {pages} {6983} (\bibinfo {year} {2015})}\BibitemShut
  {NoStop}%
\bibitem [{\citenamefont {Kelly}\ \emph {et~al.}(2015)\citenamefont {Kelly},
  \citenamefont {Barends}, \citenamefont {Fowler}, \citenamefont {Megrant},
  \citenamefont {Jeffrey}, \citenamefont {White}, \citenamefont {Sank},
  \citenamefont {Mutus}, \citenamefont {Campbell}, \citenamefont {Chen} \emph
  {et~al.}}]{kelly2015state}%
  \BibitemOpen
  \bibfield  {author} {\bibinfo {author} {\bibfnamefont {J.}~\bibnamefont
  {Kelly}}, \bibinfo {author} {\bibfnamefont {R.}~\bibnamefont {Barends}},
  \bibinfo {author} {\bibfnamefont {A.~G.}\ \bibnamefont {Fowler}}, \bibinfo
  {author} {\bibfnamefont {A.}~\bibnamefont {Megrant}}, \bibinfo {author}
  {\bibfnamefont {E.}~\bibnamefont {Jeffrey}}, \bibinfo {author} {\bibfnamefont
  {T.~C.}\ \bibnamefont {White}}, \bibinfo {author} {\bibfnamefont
  {D.}~\bibnamefont {Sank}}, \bibinfo {author} {\bibfnamefont {J.~Y.}\
  \bibnamefont {Mutus}}, \bibinfo {author} {\bibfnamefont {B.}~\bibnamefont
  {Campbell}}, \bibinfo {author} {\bibfnamefont {Y.}~\bibnamefont {Chen}},
  \emph {et~al.},\ }\href@noop {} {\bibfield  {journal} {\bibinfo  {journal}
  {Nature}\ }\textbf {\bibinfo {volume} {519}},\ \bibinfo {pages} {66}
  (\bibinfo {year} {2015})}\BibitemShut {NoStop}%
\bibitem [{\citenamefont {Lao}\ \emph {et~al.}(2019{\natexlab{a}})\citenamefont
  {Lao}, \citenamefont {van Wee}, \citenamefont {Ashraf}, \citenamefont {van
  Someren}, \citenamefont {Khammassi}, \citenamefont {Bertels},\ and\
  \citenamefont {Almudever}}]{lao2018mapping}%
  \BibitemOpen
  \bibfield  {author} {\bibinfo {author} {\bibfnamefont {L.}~\bibnamefont
  {Lao}}, \bibinfo {author} {\bibfnamefont {B.}~\bibnamefont {van Wee}},
  \bibinfo {author} {\bibfnamefont {I.}~\bibnamefont {Ashraf}}, \bibinfo
  {author} {\bibfnamefont {J.}~\bibnamefont {van Someren}}, \bibinfo {author}
  {\bibfnamefont {N.}~\bibnamefont {Khammassi}}, \bibinfo {author}
  {\bibfnamefont {K.}~\bibnamefont {Bertels}}, \ and\ \bibinfo {author}
  {\bibfnamefont {C.}~\bibnamefont {Almudever}},\ }\href@noop {} {\bibfield
  {journal} {\bibinfo  {journal} {Quantum Science and Technology}\ }\textbf
  {\bibinfo {volume} {4}},\ \bibinfo {pages} {015005} (\bibinfo {year}
  {2019}{\natexlab{a}})}\BibitemShut {NoStop}%
\bibitem [{\citenamefont {Li}\ \emph {et~al.}(2019)\citenamefont {Li},
  \citenamefont {Ding},\ and\ \citenamefont {Xie}}]{li2019tackling}%
  \BibitemOpen
  \bibfield  {author} {\bibinfo {author} {\bibfnamefont {G.}~\bibnamefont
  {Li}}, \bibinfo {author} {\bibfnamefont {Y.}~\bibnamefont {Ding}}, \ and\
  \bibinfo {author} {\bibfnamefont {Y.}~\bibnamefont {Xie}},\ }in\ \href@noop
  {} {\emph {\bibinfo {booktitle} {Proceedings of the Twenty-Fourth
  International Conference on Architectural Support for Programming Languages
  and Operating Systems}}}\ (\bibinfo {organization} {ACM},\ \bibinfo {year}
  {2019})\ pp.\ \bibinfo {pages} {1001--1014}\BibitemShut {NoStop}%
\bibitem [{\citenamefont {Tannu}\ and\ \citenamefont
  {Qureshi}(2019)}]{tannu2019not}%
  \BibitemOpen
  \bibfield  {author} {\bibinfo {author} {\bibfnamefont {S.~S.}\ \bibnamefont
  {Tannu}}\ and\ \bibinfo {author} {\bibfnamefont {M.~K.}\ \bibnamefont
  {Qureshi}},\ }in\ \href@noop {} {\emph {\bibinfo {booktitle} {Proceedings of
  the Twenty-Fourth International Conference on Architectural Support for
  Programming Languages and Operating Systems}}}\ (\bibinfo {organization}
  {ACM},\ \bibinfo {year} {2019})\ pp.\ \bibinfo {pages} {987--999}\BibitemShut
  {NoStop}%
\bibitem [{\citenamefont {Shi}\ \emph {et~al.}(2019)\citenamefont {Shi},
  \citenamefont {Leung}, \citenamefont {Gokhale}, \citenamefont {Rossi},
  \citenamefont {Schuster}, \citenamefont {Hoffmann},\ and\ \citenamefont
  {Chong}}]{shi2019optimized}%
  \BibitemOpen
  \bibfield  {author} {\bibinfo {author} {\bibfnamefont {Y.}~\bibnamefont
  {Shi}}, \bibinfo {author} {\bibfnamefont {N.}~\bibnamefont {Leung}}, \bibinfo
  {author} {\bibfnamefont {P.}~\bibnamefont {Gokhale}}, \bibinfo {author}
  {\bibfnamefont {Z.}~\bibnamefont {Rossi}}, \bibinfo {author} {\bibfnamefont
  {D.~I.}\ \bibnamefont {Schuster}}, \bibinfo {author} {\bibfnamefont
  {H.}~\bibnamefont {Hoffmann}}, \ and\ \bibinfo {author} {\bibfnamefont
  {F.~T.}\ \bibnamefont {Chong}},\ }in\ \href@noop {} {\emph {\bibinfo
  {booktitle} {Proceedings of the Twenty-Fourth International Conference on
  Architectural Support for Programming Languages and Operating Systems}}}\
  (\bibinfo {organization} {ACM},\ \bibinfo {year} {2019})\ pp.\ \bibinfo
  {pages} {1031--1044}\BibitemShut {NoStop}%
\bibitem [{\citenamefont {Lao}\ \emph {et~al.}(2019{\natexlab{b}})\citenamefont
  {Lao}, \citenamefont {Manzano}, \citenamefont {van Someren}, \citenamefont
  {Ashraf},\ and\ \citenamefont {Almudever}}]{lao2019mapping}%
  \BibitemOpen
  \bibfield  {author} {\bibinfo {author} {\bibfnamefont {L.}~\bibnamefont
  {Lao}}, \bibinfo {author} {\bibfnamefont {D.~M.}\ \bibnamefont {Manzano}},
  \bibinfo {author} {\bibfnamefont {H.}~\bibnamefont {van Someren}}, \bibinfo
  {author} {\bibfnamefont {I.}~\bibnamefont {Ashraf}}, \ and\ \bibinfo {author}
  {\bibfnamefont {C.~G.}\ \bibnamefont {Almudever}},\ }\href@noop {} {\bibfield
   {journal} {\bibinfo  {journal} {arXiv:1908.04226}\ } (\bibinfo {year}
  {2019}{\natexlab{b}})}\BibitemShut {NoStop}%
\bibitem [{\citenamefont {Gottesman}(1998)}]{gottesman1998heisenberg}%
  \BibitemOpen
  \bibfield  {author} {\bibinfo {author} {\bibfnamefont {D.}~\bibnamefont
  {Gottesman}},\ }\href@noop {} {\bibfield  {journal} {\bibinfo  {journal}
  {arXiv:9807006}\ } (\bibinfo {year} {1998})}\BibitemShut {NoStop}%
\bibitem [{\citenamefont {Fowler}(2015)}]{fowler2015minimum}%
  \BibitemOpen
  \bibfield  {author} {\bibinfo {author} {\bibfnamefont {A.~G.}\ \bibnamefont
  {Fowler}},\ }\href@noop {} {\bibfield  {journal} {\bibinfo  {journal}
  {Quantum Information \& Computation}\ }\textbf {\bibinfo {volume} {15}},\
  \bibinfo {pages} {145} (\bibinfo {year} {2015})}\BibitemShut {NoStop}%
\bibitem [{\citenamefont {Duclos-Cianci}\ and\ \citenamefont
  {Poulin}(2010)}]{duclos2010fast}%
  \BibitemOpen
  \bibfield  {author} {\bibinfo {author} {\bibfnamefont {G.}~\bibnamefont
  {Duclos-Cianci}}\ and\ \bibinfo {author} {\bibfnamefont {D.}~\bibnamefont
  {Poulin}},\ }\href@noop {} {\bibfield  {journal} {\bibinfo  {journal} {Phys.
  Rev. Lett.}\ }\textbf {\bibinfo {volume} {104}},\ \bibinfo {pages} {050504}
  (\bibinfo {year} {2010})}\BibitemShut {NoStop}%
\bibitem [{\citenamefont {Krastanov}\ and\ \citenamefont
  {Jiang}(2017)}]{krastanov2017deep}%
  \BibitemOpen
  \bibfield  {author} {\bibinfo {author} {\bibfnamefont {S.}~\bibnamefont
  {Krastanov}}\ and\ \bibinfo {author} {\bibfnamefont {L.}~\bibnamefont
  {Jiang}},\ }\href@noop {} {\bibfield  {journal} {\bibinfo  {journal}
  {Scientific reports}\ }\textbf {\bibinfo {volume} {7}},\ \bibinfo {pages}
  {11003} (\bibinfo {year} {2017})}\BibitemShut {NoStop}%
\bibitem [{\citenamefont {Varsamopoulos}\ \emph {et~al.}(2017)\citenamefont
  {Varsamopoulos}, \citenamefont {Criger},\ and\ \citenamefont
  {Bertels}}]{varsamopoulos2017decoding}%
  \BibitemOpen
  \bibfield  {author} {\bibinfo {author} {\bibfnamefont {S.}~\bibnamefont
  {Varsamopoulos}}, \bibinfo {author} {\bibfnamefont {B.}~\bibnamefont
  {Criger}}, \ and\ \bibinfo {author} {\bibfnamefont {K.}~\bibnamefont
  {Bertels}},\ }\href@noop {} {\bibfield  {journal} {\bibinfo  {journal}
  {Quantum Science and Technology}\ }\textbf {\bibinfo {volume} {3}},\ \bibinfo
  {pages} {015004} (\bibinfo {year} {2017})}\BibitemShut {NoStop}%
\bibitem [{\citenamefont {Baireuther}\ \emph {et~al.}(2018)\citenamefont
  {Baireuther}, \citenamefont {O'Brien}, \citenamefont {Tarasinski},\ and\
  \citenamefont {Beenakker}}]{baireuther2018machine}%
  \BibitemOpen
  \bibfield  {author} {\bibinfo {author} {\bibfnamefont {P.}~\bibnamefont
  {Baireuther}}, \bibinfo {author} {\bibfnamefont {T.~E.}\ \bibnamefont
  {O'Brien}}, \bibinfo {author} {\bibfnamefont {B.}~\bibnamefont {Tarasinski}},
  \ and\ \bibinfo {author} {\bibfnamefont {C.~W.}\ \bibnamefont {Beenakker}},\
  }\href@noop {} {\bibfield  {journal} {\bibinfo  {journal} {Quantum}\ }\textbf
  {\bibinfo {volume} {2}},\ \bibinfo {pages} {48} (\bibinfo {year}
  {2018})}\BibitemShut {NoStop}%
\bibitem [{\citenamefont {Ni}(2018)}]{ni2018neural}%
  \BibitemOpen
  \bibfield  {author} {\bibinfo {author} {\bibfnamefont {X.}~\bibnamefont
  {Ni}},\ }\href@noop {} {\bibfield  {journal} {\bibinfo  {journal}
  {arXiv:1809.06640}\ } (\bibinfo {year} {2018})}\BibitemShut {NoStop}%
\bibitem [{\citenamefont {Horsman}\ \emph {et~al.}(2012)\citenamefont
  {Horsman}, \citenamefont {Fowler}, \citenamefont {Devitt},\ and\
  \citenamefont {Van~Meter}}]{horsman2012surface}%
  \BibitemOpen
  \bibfield  {author} {\bibinfo {author} {\bibfnamefont {C.}~\bibnamefont
  {Horsman}}, \bibinfo {author} {\bibfnamefont {A.~G.}\ \bibnamefont {Fowler}},
  \bibinfo {author} {\bibfnamefont {S.}~\bibnamefont {Devitt}}, \ and\ \bibinfo
  {author} {\bibfnamefont {R.}~\bibnamefont {Van~Meter}},\ }\href@noop {}
  {\bibfield  {journal} {\bibinfo  {journal} {New Journal of Physics}\ }\textbf
  {\bibinfo {volume} {14}},\ \bibinfo {pages} {123011} (\bibinfo {year}
  {2012})}\BibitemShut {NoStop}%
\bibitem [{\citenamefont {Landahl}\ and\ \citenamefont
  {Ryan-Anderson}(2014)}]{landahl2014quantum}%
  \BibitemOpen
  \bibfield  {author} {\bibinfo {author} {\bibfnamefont {A.~J.}\ \bibnamefont
  {Landahl}}\ and\ \bibinfo {author} {\bibfnamefont {C.}~\bibnamefont
  {Ryan-Anderson}},\ }\href@noop {} {\bibfield  {journal} {\bibinfo  {journal}
  {arXiv:1407.5103}\ } (\bibinfo {year} {2014})}\BibitemShut {NoStop}%
\bibitem [{\citenamefont {Bomb{\'\i}n}\ and\ \citenamefont
  {Martin-Delgado}(2009)}]{bombin2009quantum}%
  \BibitemOpen
  \bibfield  {author} {\bibinfo {author} {\bibfnamefont {H.}~\bibnamefont
  {Bomb{\'\i}n}}\ and\ \bibinfo {author} {\bibfnamefont {M.~A.}\ \bibnamefont
  {Martin-Delgado}},\ }\href@noop {} {\bibfield  {journal} {\bibinfo  {journal}
  {Journal of Physics A: Mathematical and Theoretical}\ }\textbf {\bibinfo
  {volume} {42}},\ \bibinfo {pages} {095302} (\bibinfo {year}
  {2009})}\BibitemShut {NoStop}%
\bibitem [{\citenamefont {Vuillot}\ \emph {et~al.}(2019)\citenamefont
  {Vuillot}, \citenamefont {Lao}, \citenamefont {Criger}, \citenamefont
  {Garc{\'\i}a~Almud{\'e}ver}, \citenamefont {Bertels},\ and\ \citenamefont
  {Terhal}}]{vuillot2018code}%
  \BibitemOpen
  \bibfield  {author} {\bibinfo {author} {\bibfnamefont {C.}~\bibnamefont
  {Vuillot}}, \bibinfo {author} {\bibfnamefont {L.}~\bibnamefont {Lao}},
  \bibinfo {author} {\bibfnamefont {B.}~\bibnamefont {Criger}}, \bibinfo
  {author} {\bibfnamefont {C.}~\bibnamefont {Garc{\'\i}a~Almud{\'e}ver}},
  \bibinfo {author} {\bibfnamefont {K.}~\bibnamefont {Bertels}}, \ and\
  \bibinfo {author} {\bibfnamefont {B.~M.}\ \bibnamefont {Terhal}},\
  }\href@noop {} {\bibfield  {journal} {\bibinfo  {journal} {New Journal of
  Physics}\ }\textbf {\bibinfo {volume} {21}},\ \bibinfo {pages} {033028}
  (\bibinfo {year} {2019})}\BibitemShut {NoStop}%
\bibitem [{\citenamefont {Abadi}\ \emph {et~al.}(2016)\citenamefont {Abadi},
  \citenamefont {Barham}, \citenamefont {Chen}, \citenamefont {Chen},
  \citenamefont {Davis}, \citenamefont {Dean}, \citenamefont {Devin},
  \citenamefont {Ghemawat}, \citenamefont {Irving}, \citenamefont {Isard} \emph
  {et~al.}}]{abadi2016tensorflow}%
  \BibitemOpen
  \bibfield  {author} {\bibinfo {author} {\bibfnamefont {M.}~\bibnamefont
  {Abadi}}, \bibinfo {author} {\bibfnamefont {P.}~\bibnamefont {Barham}},
  \bibinfo {author} {\bibfnamefont {J.}~\bibnamefont {Chen}}, \bibinfo {author}
  {\bibfnamefont {Z.}~\bibnamefont {Chen}}, \bibinfo {author} {\bibfnamefont
  {A.}~\bibnamefont {Davis}}, \bibinfo {author} {\bibfnamefont
  {J.}~\bibnamefont {Dean}}, \bibinfo {author} {\bibfnamefont {M.}~\bibnamefont
  {Devin}}, \bibinfo {author} {\bibfnamefont {S.}~\bibnamefont {Ghemawat}},
  \bibinfo {author} {\bibfnamefont {G.}~\bibnamefont {Irving}}, \bibinfo
  {author} {\bibfnamefont {M.}~\bibnamefont {Isard}},  \emph {et~al.},\ }in\
  \href@noop {} {\emph {\bibinfo {booktitle} {12th $\{$USENIX$\}$ Symposium on
  Operating Systems Design and Implementation ($\{$OSDI$\}$ 16)}}}\ (\bibinfo
  {year} {2016})\ pp.\ \bibinfo {pages} {265--283}\BibitemShut {NoStop}%
\end{thebibliography}%
\end{document}